\documentclass[fleqn,usenatbib]{mnras}

\usepackage{newtxtext,newtxmath}

\usepackage[T1]{fontenc}
\usepackage{ulem}

\DeclareRobustCommand{\VAN}[3]{#2}
\let\VANthebibliography\thebibliography
\def\thebibliography{\DeclareRobustCommand{\VAN}[3]{##3}\VANthebibliography}

\usepackage{graphicx}	

\def \twomass {2MASS J00524508-7228437}
\def \scwd {CXOU J005245.0-722844} 
\def \bexrb {BeXRB}
\def \bexrbs {BeXRBs}
\def \scubed {S-CUBED}
\def \swift {Swift}

\title [CXOU J005245.0-722844]{\scwd: Discovery of a Be Star / White Dwarf binary system in the SMC via a very fast, super-Eddington X-ray outburst event }

\author[T.~M. Gaudin et al.]{T.~M. Gaudin$^{1}$\thanks{E-mail: tmg6006@psu.edu (TMG)},
M.~J. Coe$^{2}$,
J.~A. Kennea$^{1}$,
I.~M. Monageng$^{3,6}$,
D.~A.~H. Buckley$^{3,6}$,
A. Udalski$^{4}$,
P.~A. Evans$^{5}$
\\
$^{1}$Department of Astronomy and Astrophysics, The Pennsylvania State University, 525 Davey Lab, University Park, PA 16802, USA\\
$^{2}$Physics \& Astronomy, The University of Southampton, SO17 1BJ, UK\\
$^{3}$South African Astronomical Observatory, P.O Box 9, Observatory, 7935, Cape Town, South Africa\\
$^{4}$Astronomical Observatory, University of Warsaw, Al. Ujazdowskie 4, 00-478 Warszawa, Poland \\
$^{5}$University of Leicester, X-ray and Observational Astronomy Research Group, School of Physics \& Astronomy, University Road, Leicester LE1 7RH, UK
\\
$^{6}$Department of Astronomy, University of Cape Town, Private Bag X3, 7701 Rondebosch, South Africa
}
\date{Accepted 2021 September 9. Received 2021 August 31; in original form 2021 July 29}

\pubyear{2021}

\begin{document}
\label{firstpage}
\pagerange{\pageref{firstpage}--\pageref{lastpage}}
\maketitle

\begin{abstract}

\scwd{} is an X-ray source in the Small Magellanic Cloud (SMC) that has long been known as a Be/X-ray binary (BeXRB) star, containing an OBe main sequence star and a compact object. In this paper, we report on a new very fast X-ray outburst from \scwd{}. X-ray observations taken by Swift constrain the duration of the outburst to less than 16 days and find that the source reached super-Eddington X-ray luminosities during the initial phases of the eruption. The XRT spectrum of \scwd{} during this outburst reveals a super-soft X-ray source, best fit by an absorbed thermal blackbody model. Optical and Ultraviolet follow-up observations from the Optical Gravitational Lensing Experiment (OGLE), Asteroid Terrestrial-impact Last Alert System (ATLAS), and Swift identify a brief $\sim$0.5 magnitude optical burst coincident with the X-ray outburst that lasted for less than 7 days. Optical photometry additionally identifies the orbital period of the system to be 17.55 days and identifies a shortening of the period to 17.14 days in the years leading up to the outburst. Optical spectroscopy from the Southern African Large Telescope (SALT) confirms that the optical companion is an early-type OBe star. We conclude from our observations that the compact object in this system is a white dwarf (WD), making this the seventh candidate Be/WD X-ray binary. The X-ray outburst is found to be the result of a very-fast, ultra-luminous nova similar to the outburst of MAXI J0158-744.

\end{abstract}




\section{Introduction}

Be/X-ray binaries (\bexrbs) are a type of high mass X-ray binary (HMXB) system containing a main sequence O or B type star in orbit with a compact object, typically a Neutron Star (NS), in a moderately eccentric orbit ($e\sim0.3-0.5$). The OB star is characterized by the presence of Balmer-series Hydrogen emission lines in its optical spectrum from which the spectral class Be is derived. These emission lines are the signature of a dense, circumstellar wind that forms a geometrically-thin ``decretion disk" around the main sequence star and fuels X-ray emission via accretion onto the compact object (see \citealt{Reig11} for a review).

BeXRBs are commonly-observed, making up over half of the galactic population of HMXBs \citep{2023Neumann}. These systems are also numerous in the Small Magellanic Cloud (SMC), a neighboring dwarf galaxy of the Milky Way \citep{CK2015, 2017Yang}. Due to a recent period of increased star formation \citep{2004Harris, 2014Rezaeikh}, the HMXB population of the SMC is much larger than is expected for a low-mass dwarf galaxy. Almost all HXMBs (SMC X-1; \citealt{1997Li}) in this population are observed to be BeXRBs, and the total number of BeXRBs is comparable to the size of the galactic population. The study of this large population of BeXRBs is the primary motivation for the Swift SMC Survey (S-CUBED; \citealt{kennea2018}). S-CUBED is a weekly survey, ongoing since 2016, that aims to identify new BeXRBs and characterize their high-energy emission behavior. This survey utilizes the X-ray Telescope (XRT; \citealt{Roming05}) and UV/Optical Telescope (UVOT; \citealt{burrows05}) of the Swift Observatory to perform tiled observations of 142 overlapping tiles for 60s each in order to obtain spatially-continuous observations of the entire SMC. Each tile is observed with XRT in Photon Counting (PC) mode and UVOT observing the \textit{uvw1}-band. S-CUBED data has been used to identify several new BeXRBs \citep{Kennea2020, Monageng2019, 2024Gaudin} and to observe notable outbursts from known systems (e.g. SMC X-3; \citealt{Townsend2017}).

While the compact object in a BeXRB system is typically a NS, there have been cases where either a WD (e.g. XMMU J010147.5-715550; \citealt{sturm2012}) or a black hole \citep{2014Casares} has been found in orbit around a Be star. These cases are rare, and only 6 candidate Be/White Dwarf (henceforth BeWD) systems are known to exist. Of the six candidate BeWD systems that have been identified, four have been found in the SMC \citep{2023Zhu}, and two of the four have been discovered via S-CUBED observations \citep{coe2020b, 2021Kennea}. The reason for a lack of observed BeWD systems remains an open question. It has been suggested by population synthesis studies that BeWD systems should be significantly more prevalent than BeXRB systems with a NS compact object \citep{raguzova2001, 2023Zhu}. The results of \citet{raguzova2001} predict an occurrence rate of BeWDs over 7 times as high as BeXRBs containing a NS. However, observational evidence for this large population of hidden BeWDs remains elusive.

Observational selection bias has been cited \citep{2012Sturm, 2021Kennea} as a potential reason for the lack of detected BeWDs. BeXRBs are transient accretion-powered systems, and X-ray outbursts are thought to be the result of interactions between the Be circumstellar disk and the compact object \citep{1986Stella}. When the compact object is a NS, the luminosity of an accretion-powered X-ray outburst is $L_X \sim 10^{36} - 10^{38}$ ergs s$^{-1}$ \citep{okazaki2001}. However, WD accretion is predicted to be much weaker, producing hard X-ray emission luminosity of $L_X \sim 10^{29} - 10^{33}$ ergs s$^{-1}$ \citep{1989Waters}, which is below the detection threshold of wide field X-ray telescopes such as the Monitor of All-sky X-ray Image (MAXI) telescope or the Swift Burst Alert Telescope (BAT) for sources in the Magellanic Clouds. When BeWD systems are detected, the X-ray spectrum is found to be that of a super-soft X-ray source (SSS) \citep{coe2020b} with most X-ray emission detected below 2 keV. SSS are a class of accreting WDs that are experiencing a period of stable nuclear burning on their surface \citep{kahabka2006}. During the SSS phase, the WD can produce X-ray luminosities of $L_X \sim 10^{36} - 10^{38}$ ergs s$^{-1}$ \citep{1997Kahabka}, making them far easier to detect by X-ray telescope observations.

In rare cases, SSS can reach abnormally large luminosities that are significantly higher than the Eddington Luminosity of a 1 M$_\odot$ WD ($L_X \geq 10^{39}$ ergs s$^{-1}$). These systems, designated as ultra-luminous supersoft sources (ULSs) \citep{2008Liu}, have no obvious physical origin. Some \citep{2008Liu} have suggested that intermediate-mass black holes may be responsible for this rare class of objects,and others favor a scenario in which a stellar mass black hole is accreting at a super-critical rate \citep{2016Urquhart}. In extremely rare cases, a nuclear burning WD will briefly reach the luminosity regime of an ULS. This has happened only once, in the SMC BeWD system MAXI J0158-744 \citep{2012Li, 2013Morii}. MAXI J0158-744 was discovered via a brief soft X-ray outburst that reached a maximum luminosity of $L_X = 2 \times 10^{40}$ ergs s$^{-1}$ \citep{2013Morii} and lasted for only weeks \citep{2012Li}. This eruption is the first example of a nova eruption occurring in a BeWD system during which a shell of accreted Hydrogen on the surface of the WD produced a period of runaway nuclear fusion (see \citealt{2022Page} for a recent review of novae) that resulted in a bright outburst. The importance of BeWD systems to ULS systems and the ultra-luminous X-ray source (ULX) phenomenon is still an area of active research. In order to better understand the connection, more systems must be observed.

In this paper, we report on the recent super-Eddington outburst of \scwd{}, a new BeXRB in the SMC that becomes the second BeWD to reach the ULS regime. This new system was observed by combining Swift XRT and UVOT observations, archival data from the Optical Gravitational Lensing Experiment (OGLE) and the Asteroid Terrestrial-impact Last Alert System (ATLAS), and Southern African Large Telescope (SALT) spectroscopy. Observational data can be combined to show that the compact object in this system is indeed a WD, making this the fifth candidate BeWD system identified in the SMC. We also demonstrate that the super-luminous outburst is likely produced by the beginning phases of a nova eruption on the surface of the compact object.

\section{Observations}

\subsection{Discovery of \scwd}

\begin{figure*}
	\includegraphics[scale=0.48]{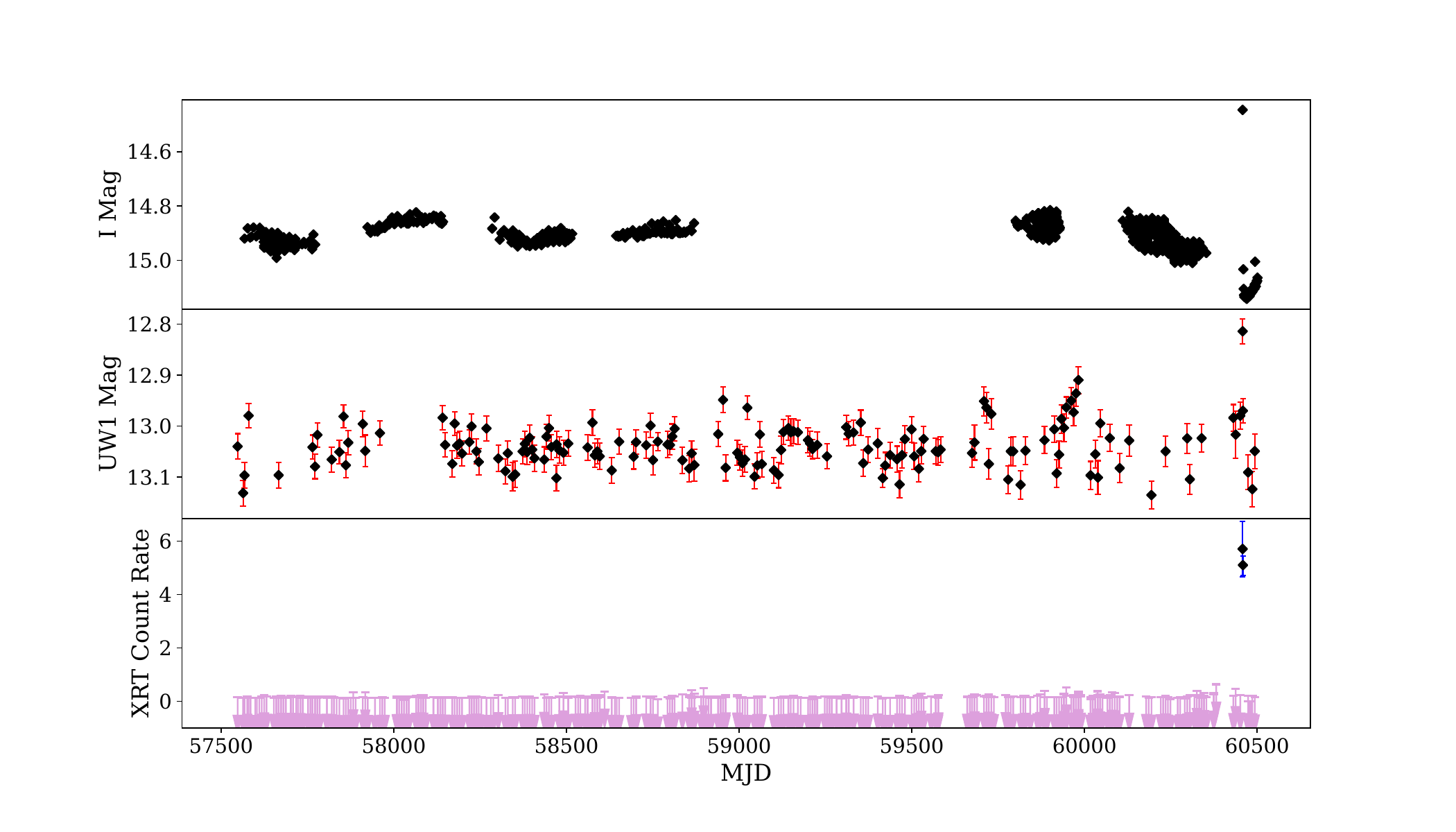}
    \caption{Multi-wavelength light curve of \scwd{} emission behavior containing I-band data taken by OGLE, \textit{uvw1}-band data taken by Swift UVOT, and 0.3-10 keV data taken by Swift XRT. In the bottom panel, black points represent observations during which \scwd{} was detected by Swift XRT. Purple arrows represent upper limits on the count rate during observations where \scwd{} was not able to be detected.}
    \label{fig:scubed_lc}
\end{figure*}

\scwd{} was first cataloged as part of the \textit{Chandra} survey of the Small Magellanic Cloud during a search for optical counterparts to SMC X-ray sources \citep{2009Antoniou}. Since its discovery, the source has been in SMC BeXRB catalogs such as \citet{Haberl2016}, but little has been known about the source. The first detection of an X-ray flare came from the Einstein Probe on 2024 May 27 at 8:41 UTC (JD2460457.862)\citep{2024Yang}. The initial Einstein Probe data was used to localize the flare to the position of \scwd{} and to note the soft spectrum of the X-ray emission that was best fit by an absorbed blackbody. About 14 hours later, at 22:29 UTC, the position of the flare was further localized by a weekly \scubed{} observation \citep{2024Kennea}, confirming that \scwd{} was indeed the source of the outburst. The short 55s observation was additionally used to verify the soft nature of the X-ray spectrum and posit that the compact object in this system was indeed a WD.

The \scubed{} localization of \scwd{} is the most accurate X-ray position of the flare, localizing the source to RA(J2000) = 00$^h$ 52$^m$ 45.18$^s$, Dec(J2000) = -72$^\circ$ 28$'$ 44.6$"$ with an estimated uncertainty region of radius 2.4 arcseconds. The optical counterpart of \scwd{} was found by \citet{2024Jaisawal} to be \twomass{} at a position of RA(J2000) = 00$^h$ 52$^m$ 45.09$^s$, Dec(J2000) = -72$^\circ$ 28$'$ 43.73$"$, separated from the X-ray source position by less than a single arcsecond. \twomass{} was first reported to be an early-type OBe star in 2009 when the source was discovered, with a spectral type of O9Ve-B0Ve \citep{2009Antoniou}. \citet{2024Jaisawal} reported on the observation of \scwd{} by NICER, finding a similar soft spectrum with a much fainter luminosity than the detections of Einstein Probe and \scubed{}. NICER observations provide the first evidence of an OVIII absorption edge at 0.871 keV, further strengthening the case for a WD compact object in the system. 

\begin{figure}
	\includegraphics[scale=0.4]{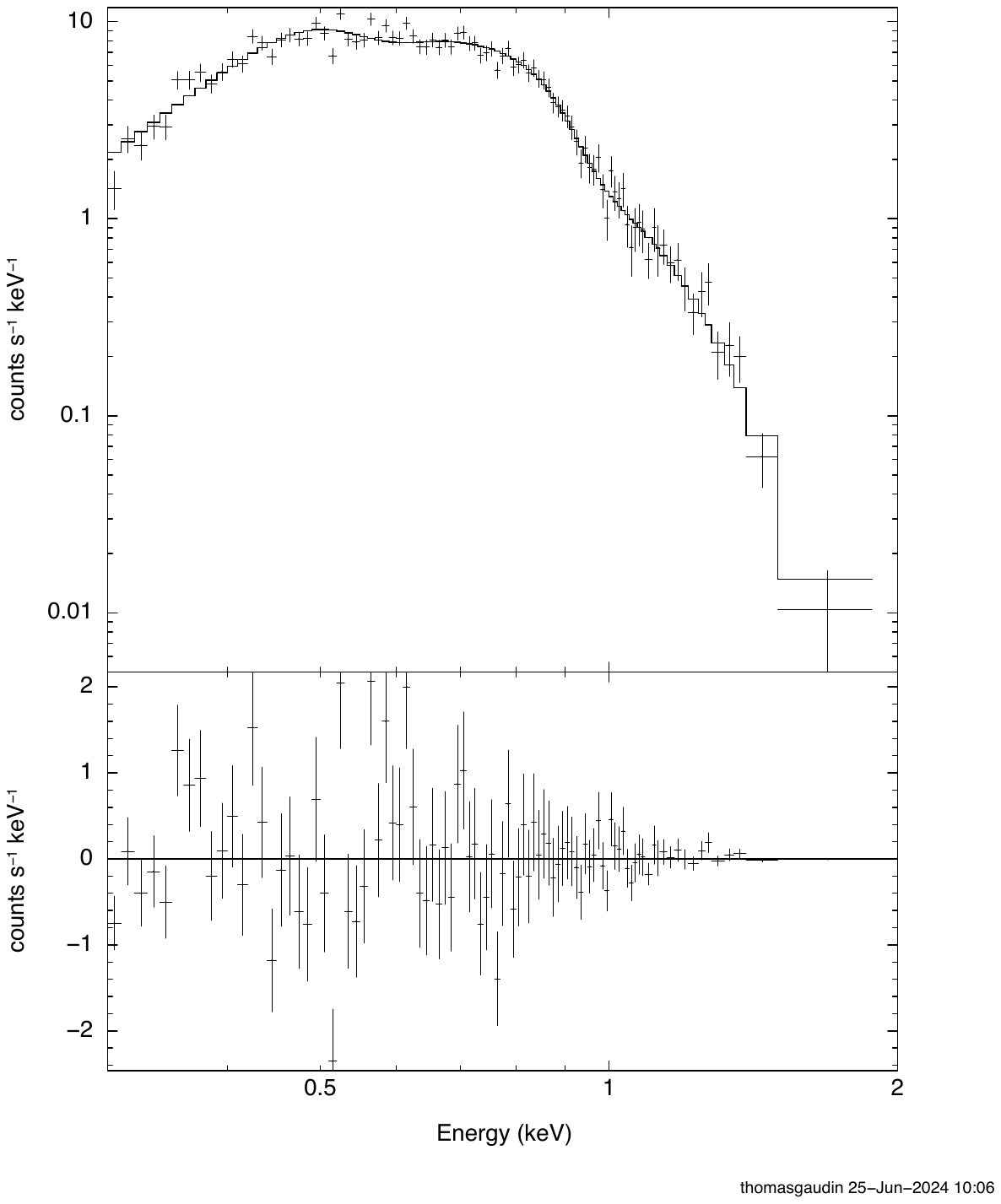}
    \caption{X-ray spectrum of \scwd{} taken during a WT mode observation on 2024 May 28 including the best-fitting absorbed blackbody model containing 2 absorption edges at 0.385 keV and 0.896 keV. The spectrum is soft with little emission detected above 2 keV.}
    \label{fig:spectrum}
\end{figure}

\subsection{\swift{} Observations}

\subsubsection{XRT Observations}

\scwd{} has rarely been detected by Swift during the short, weekly observations of the S-CUBED survey. A full XRT light curve is plotted for \scwd{} in the 3rd panel of Figure \ref{fig:scubed_lc}. The system has been observed 356 times by the survey, but prior to its recent outburst, there were no detections. The recent outburst was first detected by S-CUBED on 2024 May 27, when \scwd{} was detected at a count rate of $5.71^{+1.04}_{-1.04}$ counts s$^{-1}$. A target of opportunity (TOO) observation was scheduled for 2024 May 28 to follow-up on the detected outburst. This TOO observation was taken in window timing (WT) mode, and it revealed that the source was still in a state of outburst at a count rate of $5.10^{+0.35}_{-0.39}$ counts s$^{-1}$. After this observation, no Swift observations could happen because \scwd{} was constrained by the limb of the Earth for 9 days. By the time the source was visible again by Swift on 2024 June 12, the source had faded below the detection threshold for the short exposures of \scubed{}. In four deeper TOO observations taken between 2024 June 13 and 2024 June 19, this source was also unable to be detected, indicating that the outburst had indeed concluded before coming out of Swift's Earth limb constraint. 

\begin{table}
    \centering
    \begin{tabular}{cccc}
         \hline 
         \hline
         \texttt{xspec} Model & Parameter & Best-Fit Value & Units\\
         \hline
        \texttt{TBabs} & $N_H$ & $(2.30 \pm 0.475) \times 10^{21}$ & cm$^{-2}$ \\
        \texttt{edge} & $E_{e, 1}$ & $0.385 \pm 0.020$ & keV \\
        \texttt{edge} & $\tau_{max, 1}$ & $1.53 \pm 0.43$ & - \\
        \texttt{edge} & $E_{e, 2}$ & $0.896 \pm 0.011$ & keV \\
        \texttt{edge} & $\tau_{max, 2}$ & $1.08 \pm 0.14$ & - \\
        \texttt{cflux} & $\log_{10}\left(F_{tot}\right)$ & $-8.85 \pm 0.14$ & erg cm$^{-2}$ s$^{-1}$ \\
        \texttt{bbodyrad} & $kT$ & $91.3 \pm 3.73$  & eV  
    \end{tabular}
    \caption{Table containing the best-fitting parameters from the absorbed blackbody model used to describe the spectrum of \scwd{}. Each row describes a free parameter using the model component that it is derived from, its units, and its best-fitting values to the Swift XRT spectrum taken on 2024 May 28.}
    \label{tab:x-ray best fit}
\end{table}

A 0.3-10 keV spectrum of \scwd{} was obtained for the WT mode TOO observation by using the automated XRT pipeline tools described by \citet{Evans09} and re-binned using the \texttt{grppha} software package so that each spectral bin has at least 15 counts. A typical BeXRB spectrum with a NS companion is hard and can be fit by an absorbed power law with a photon index of $\Gamma \simeq 1$. \scwd{} does not show this typical X-ray spectrum, but instead has a very soft spectrum with most of the flux being detected at energies below 2 keV. This soft X-ray spectrum is the key distinguishing feature of BeWDs that can be used to differentiate them from systems containing a NS \citep{coe2020, 2021Kennea}.

The spectrum of \scwd{} is shown in Figure \ref{fig:spectrum} and is plotted together with the line of best fit obtained via model fitting. Spectral fitting was performed by following the analysis methods of \citet{2021Kennea}, fitting an absorbed thermal blackbody to this spectrum using \texttt{xspec} \citep{1996Arnaud} which is included in the FTOOLS software package \citep{1999Blackburn}. The best-fitting parameters from model fitting are shown in Table \ref{tab:x-ray best fit}. To create the spectral model used for fitting, the \texttt{tbabs} ISM absorption model component was convolvded with the \texttt{bbodyrad} model within \texttt{xspec}. The best-fitting column density of material along the line of sight, $N_H = (2.30 \pm 0.475) \times 10^{21}$ cm$^{-2}$, is higher than the average value for the column density towards the SMC \citep{Willingale2013, kennea2018}. \scwd{} is best-fit by a thermal blackbody with temperature $kT = 91.3 \pm 3.73$ eV, corresponding to a WD mass of $\sim$1.2 M$_{\odot}$ using the maximum shock temperature derived by \citet{Mukai17}, which assumes free-fall accretion from infinity onto the surface of a WD. The parameter \texttt{cflux} is also included in the fit in order to estimate the total 0.3-10 keV flux of the source during its outburst, and the best-fitting flux is $F_{X} = 1.41^{+0.54}_{-0.38} \times 10^{-9}$ erg cm$^{-2}$ s$^{-1}$ which corresponds to an X-ray luminosity of $6.51^{+2.5}_{-1.2} \times 10^{38}$ erg s$^{-1}$ at the standard distance to the SMC of 62.44 kpc \citep{2020Graczyk}. This brightness is slightly larger than the Eddington Luminosity, $L_{edd} = 1.5 \times 10^{38}$ erg s$^{-1}$ of a 1.2 M$_{\odot}$ WD. The large luminosity implies super-Eddington accretion onto the surface of the WD during the brief outburst.

In a similar result to those of \citet{2021Kennea}, the absorbed blackbody model fit is improved by including absorption edges using the \texttt{edge} model component of \texttt{xspec}. Two absorption edges are needed to properly characterize all of the features in the spectrum and to improve the fit from a reduced $\chi^2$ of 2.17 to a reduced $\chi^2$ of 1.41. The first edge is found to be located at $E_{e,1} = 0.385 \pm 0.020$ keV, and the second edge is found at $E_{e,2} = 0.896 \pm 0.011$ keV. Given their respective locations, these are likely signatures of the 0.49 eV C VI edge and the 0.871 eV O VIII edge, respectively. Both of these edges are predicted to be found in WDs by stellar atmosphere models, and both edges have been observed in other SSS systems \citep{1997Parmar, 1998Parmar} and BeWD systems \citep{2021Kennea}.

Using the absorbed blackbody emission model and following the methods of \citet{2021Kennea}, it is possible to derive the maximum size of the emission region during outburst of \scwd{}. The normalization factor of the \texttt{bbodyrad} model component within \texttt{xspec} is given by $n = \frac{R_{em}^2}{D_{10}^2}$, where $R_{em}$ is the radius of the source emission region in units of km and $D_{10}$ is the distance to the source in units of 10 kpc. By fixing the components of the absorbed blackbody model with two absorption edges at their best-fitting values, the normalization factor can be left as a free parameter in order to determine its best-fitting value. This value for \scwd{} is found to be $n = (3.49 \pm 0.038) \times 10^{6}$, corresponding to an emission region with a radius of $R_{em} = 11648 \pm 64$ km at the standard distance to the SMC. An emission region of this size is only a few times larger than the radius of a 1.2 M$_{\odot}$ WD \citep{2005Althaus}, providing further evidence in favor of a WD compact object as the binary companion in this system. 

The flux of the source can be used to place an upper limit on the quiescent emission from \scwd{} if we assume a constant spectrum during and after the outburst. From the spectral fit presented in Table \ref{tab:x-ray best fit}, the counts to flux conversion factor is calculated to be $3.12 \times 10^{-10}$ erg cm$^{-2}$ count$^{-1}$ using the observed count rate of 4.521 counts s$^{-1}$ and the best-fitting unabsorbed flux of $F_{X} = 1.41^{+0.54}_{-0.38} \times 10^{-9}$ erg cm$^{-2}$ s$^{-1}$. By summing all of the XRT data taken during TOO observations from 2024 June 13 to 2024 June 19, we obtain a 3-$\sigma$ upper limit for the 0.3-10 keV count rate during this period. The quiescent count rate upper limit is found to be 0.0019 counts s$^{-1}$. Multiplying this count rate by the counts to flux conversion factor, we arrive at an unabsorbed flux of $F_{X} = 5.96 \times 10^{-13}$ erg cm$^{-2}$ s$^{-1}$ which corresponds to an upper limit luminosity of $L_{X} = 2.75 \times 10^{35}$ erg s$^{-1}$ at the accepted distance of the SMC. 

During a 100ks observation taken by the \textit{Chandra} X-ray Observatory in 2010 \citep{laycock2010}, \textit{Chandra} detected 12 counts over 100ks which corresponds to a flux of $F_{X} = 3.22 \times 10^{-12}$ erg cm$^{-2}$ s$^{-1}$ or a luminosity of $L_{X} = 1.48 \times 10^{35}$ erg s$^{-1}$ at the accepted distance of the SMC. This indicates agreement in quiescent luminosity between the deepest pre-outburst observation and the upper limits derived after the outburst had concluded. However, novae typically produce hard X-ray emission with a luminsoity of 10$^{33}$ - 10$^{35}$ erg s$^{-1}$ \citep{2008Mukai} for tens of days after the onset of the outburst, so the outburst could have continued to produce X-ray emission at a luminosity below the detection threshold of Swift for a longer period of time.

\subsubsection{UVOT Observations}

Since 2016, \scwd{} has been detected by UVOT in 177 \scubed{} observations and one TOO observation. For most of the duration of \scubed{}, this represents approximately weekly coverage of the source. However, the source is located at the edge of an \scubed{} UVOT tile and can occasionally fall outside of the UVOT field of view. UVOT light curves are generated by using \swift-specific functions found in FTOOLS \citep{1999Blackburn} to perform aperture photometry. An aperture with a radius of 5" was defined around the XRT position of the source and a nearby background region was defined with a radius of 8". Photometric information for each observation was then extracted within the source region using the \texttt{uvotsource} method from FTOOLS. The UVOT light curve generated using this method is shown in the center panel of Figure \ref{fig:scubed_lc}. As can be seen in the figure, the outburst is detected in the 2024 May 27 S-CUBED observation at 12.81 magnitude, which is almost 0.2 magnitudes brighter than the previous week. By the time the system was observed during the TOO observation of 2024 May 28, the system had faded back to the quiescent 12.96 magnitude. By 2024 June 12, the new source has faded to 13.09 magnitude, which is close to the faintest magnitude ever observed for the system. The \textit{uvw1}-band light curve additionally shows variability that tracks with the IR variability of the system, an expected feature of BeXRB systems that has been observed before \citep{2021Kennea}.

\begin{figure}
	\includegraphics[scale=0.36]{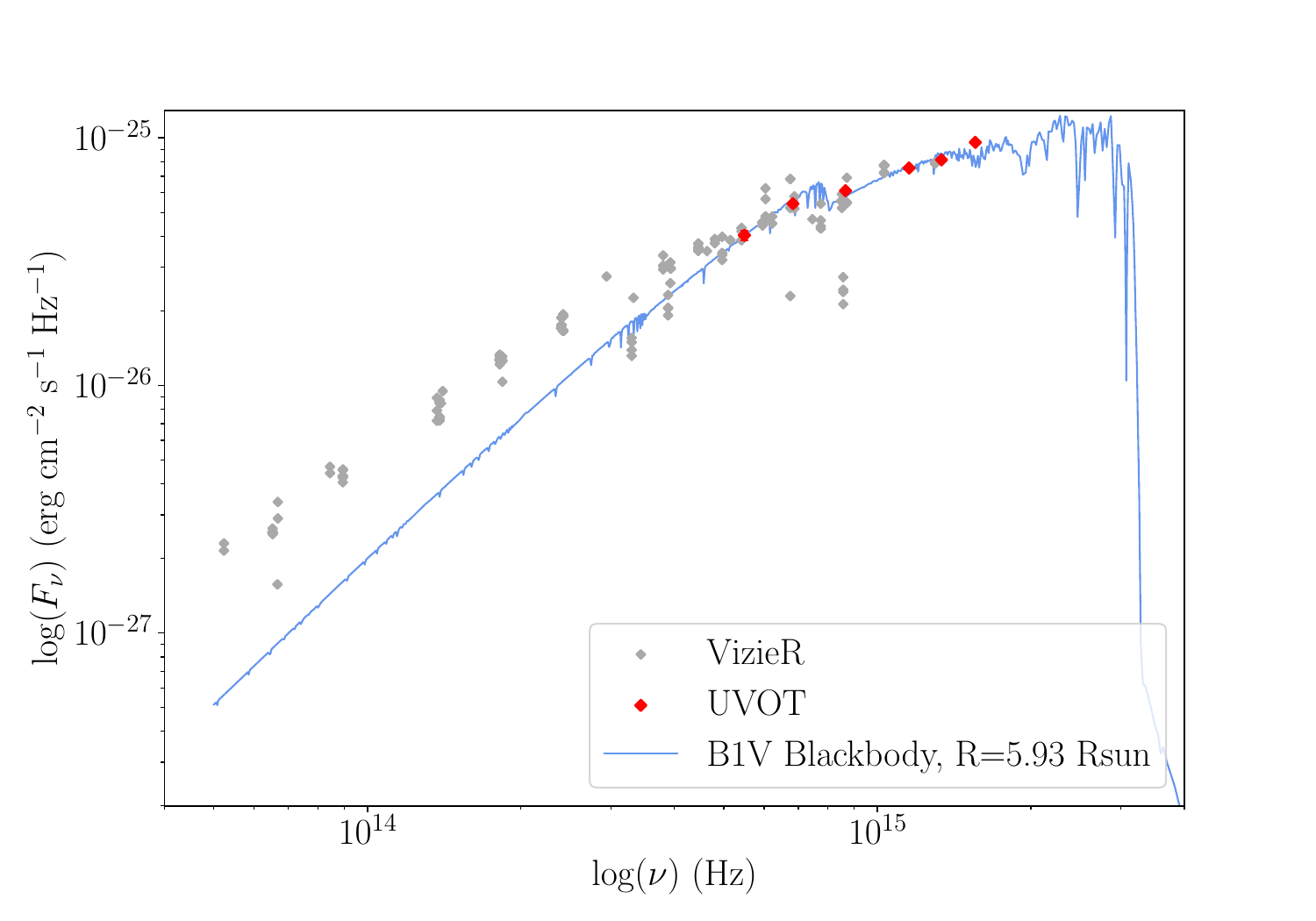}
    \caption{A multi-wavelength SED of \scwd{} created by combining Swift UVOT data with archival data retrieved from the VizieR Photometry Viewer \citep{2000Ochsenbein}. A BIV stellar spectrum is plotted alongside the SED data and scaled via use of least-squares fitting to determine the best-fit stellar radius.}
    \label{fig:sed}
\end{figure}

During the Swift TOO observation taken on 2024 May 28, \scwd{} was observed in 0x30ed mode to collect data in six different filter bands. UVOT data from this observation was combined with all archival photometric measurements of 2MASS J00524508-7228437 to create a spectral energy distribution (SED) from the infrared (IR) to the ultraviolet (UV). The archival photometric data used to create this SED were retrieved by querying the VizieR archival photometry viewer \citep{2000Ochsenbein} for all measurements made within a 1.5" radius region around the optical counterpart. Before creating the SED, all VizieR and UVOT data were unreddened using the Fitzpatrick 1999 \citep{1999Fitzpatrick} extinction law. The line-of-sight extinction towards the source was determined using the OGLE $E(V-I)$ map \citep{2021Skowron} for the Magellanic Clouds. The unreddend SED was then analyzed to check for an IR excess that is produced by the Be stellar disk. Least-squares fitting was used to fit Kurucz model spectra \citep{1993Kurucz} to the multi-wavelength SED, using the best-fitting radius to scale the spectrum to the accepted distance of $D=62.44$ kpc to the SMC \citep{2020Graczyk}. Figure \ref{fig:sed} shows the best-fitting spectrum plotted together with the UVOT and VizieR data. This figure shows that the SED is best fit to a B1V spectrum with a radius of $R_{Be} = 5.93 \pm 0.049$ R$_{\odot}$, indicating that the optical counterpart is indeed an early-type B star. In addition to fitting an early-type B star, the SED shows a clear IR excess over a standard BIV star. The strength of the IR excess of a Be star is highly time-dependent, so this SED, which combines observations taken over many years, is of limited use. However, the presence of the IR excess in many obseravtions of the time-averaged SED is a strong indicator that a circumstellar disk is present around the B-type star. While the contribution of the disk to the SED varies with time, the underlying stellar spectrum is thought to be time-invariant, so the dangers of inferring too much about the stellar disk from this SED do not apply to the confirmation of the expected spectral type via SED-fitting. 

\subsection{Optical photometry}
\subsubsection{OGLE}
The OGLE project \citep{Udalski2015} undertakes to provide long term I-band photometry with an average cadence of 1-3 days. The star \twomass{} was observed continuously for over 2 decades in the I-band with only a gap of $\sim 2.5$ year due to COVID-19 restrictions. After that gap, the source was observed with a high cadence for the following two seasons. It is identified in the OGLE catalogue as SMC719.20.104.

\begin{figure}
	\includegraphics[width=8cm,angle=-0]{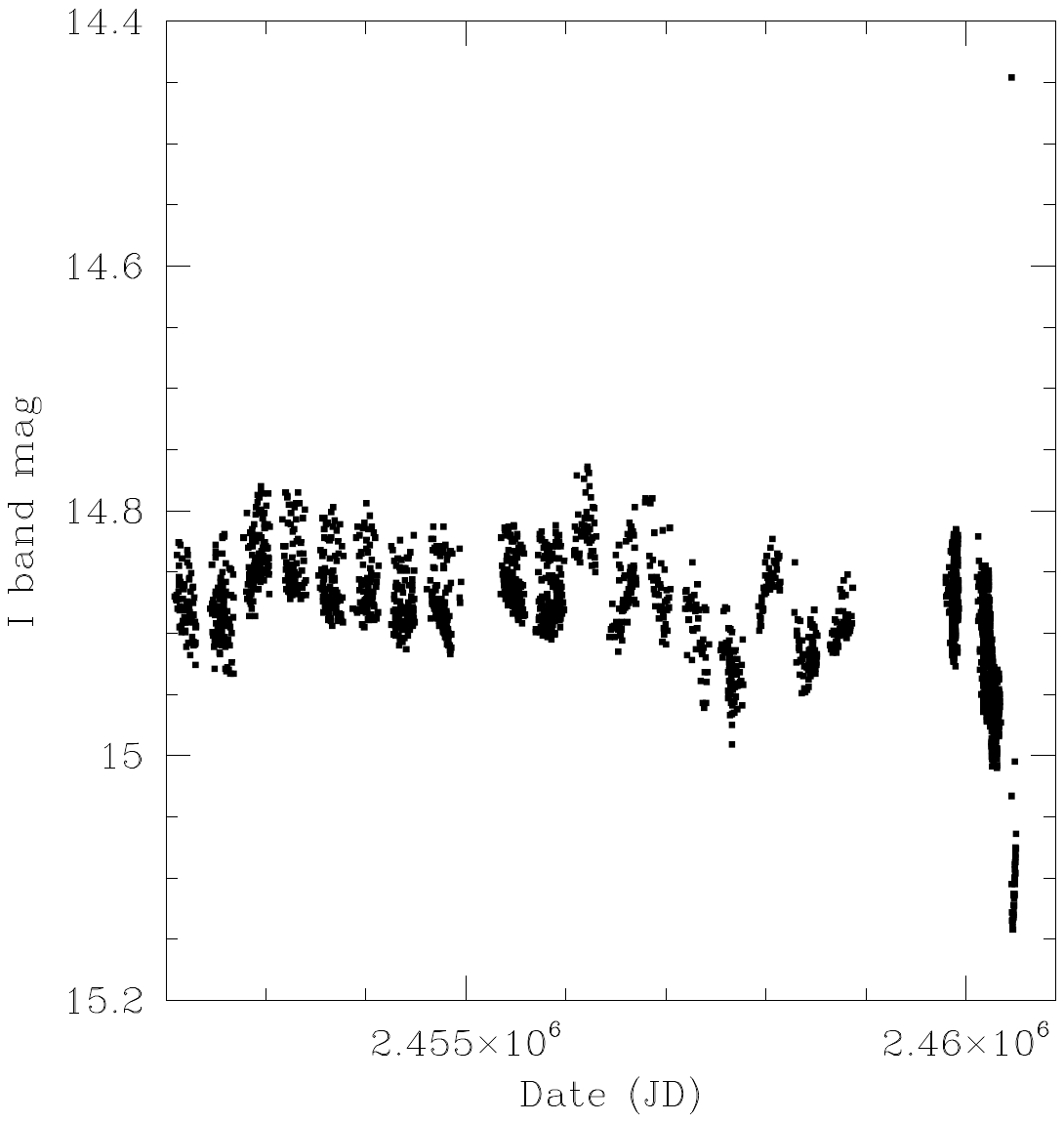}
    \caption{{OGLE III and IV observations of the optical counterpart to \scwd. Note the one outburst point in the top right hand corner of the plot followed by a very rapid fading.}}
    \label{fig:ogle}
\end{figure}

The I band data are shown in their entirety in Fig.~\ref{fig:ogle}. The overall behaviour revealed is very quiet compared to that expected for a Be star in the SMC, unusually showing no large scale fluctuations on timescales of years.  The overall colour changes seen in this system by combining the OGLE I-band with the UVOT data are discussed further in the Section~\ref{discussion} below, but again the only substantial change occurs at the time of the outburst.

OGLE detected an optical counterpart of the X-ray outburst see by the Einstein Probe at MJD 60457.62, some 6 h before the EP detection. 

\begin{figure}
    \centering
    \includegraphics[scale=0.48]{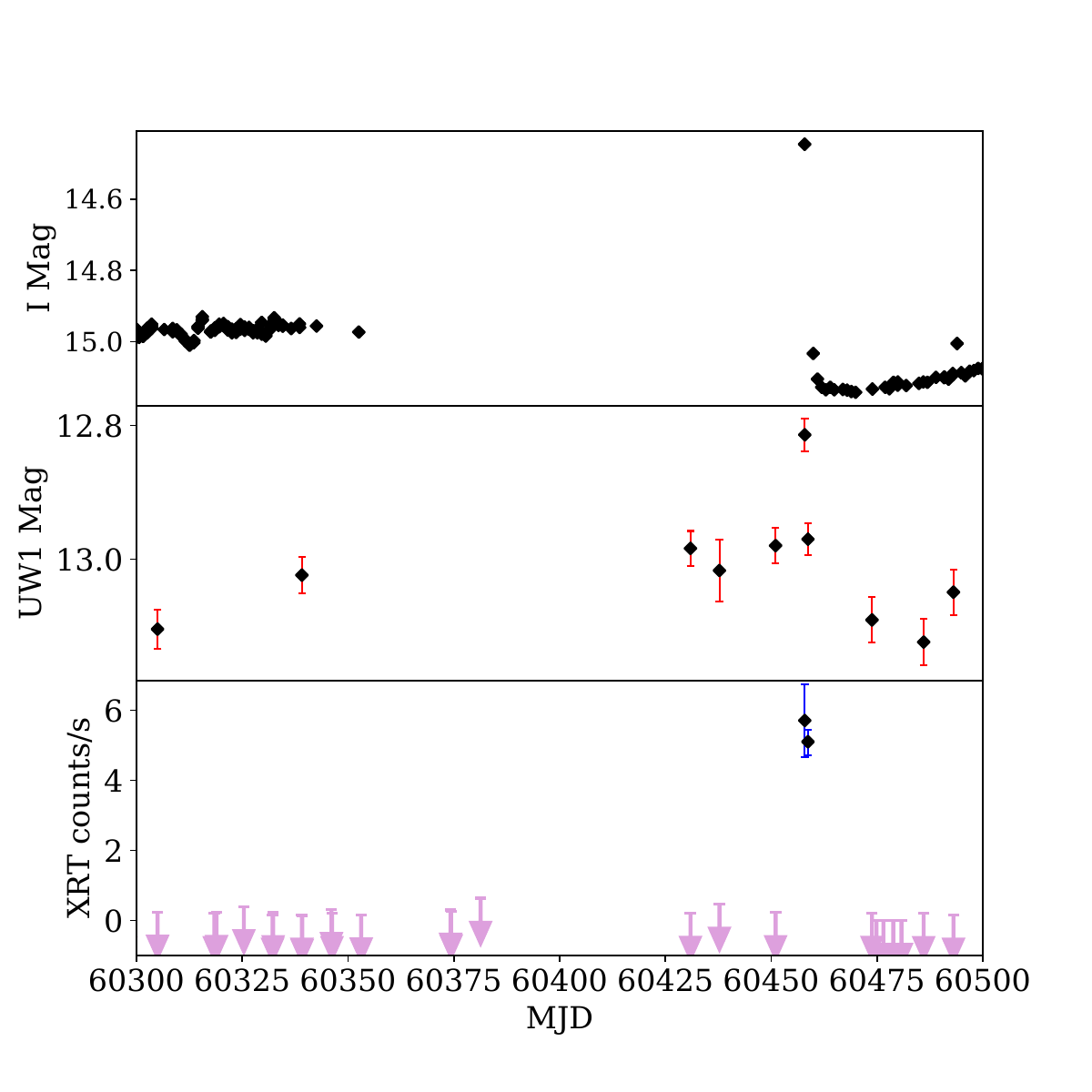}
    \caption{A multi-wavelength light curve of \scwd{} that shows the brightness of the source before, during, and after its 27 May 2024 outburst which highlights the brief IR, UV, and X-ray duration of the outburst and subsequent fading that is seen in the UV and IR. The top panel shows I-band OGLE variability. The middle panel shows \textit{uvw1}-band Swift UVOT variability. The bottom panel shows observations using Swift XRT. Black points represent observations in which \scwd{} was detected and purple arrows represent upper limits on the count rate when the source was not detected. }
    \label{fig:mw_outburst}
\end{figure}

In Figure \ref{fig:mw_outburst}, we plot higher-resolution light curves of \scwd, which for both OGLE and UVOT indicates that a phase of gradual dimming has begun, which has become fainter than the pre-outburst magnitude.

\subsubsection{ATLAS}
\begin{figure}
    \centering
    \includegraphics[scale=0.48]{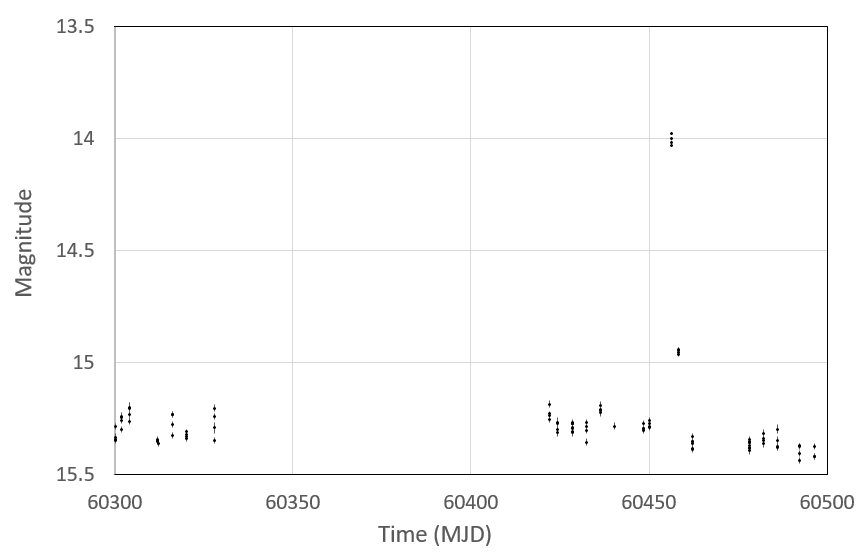}
    \caption{ATLAS -o-band photometry of \scwd{}. }
    \label{fig:ATLAS}
\end{figure}

We utilized the photometric database of the Asteroid Terrestrial-impact Last Alert System (ATLAS; \cite{2018PASP..130f4505T}) and undertook forced photometry \citep{2020PASP..132h5002S} at the position of \scwd. The light curve for the o-band photometry, covering the same zoomed-in time interval of the OGLE photometry (Figure \ref{fig:mw_outburst}), is shown in Figure \ref{fig:ATLAS}. These data show that the optical outburst was in progress from MJD 60456.370, $\sim$1  day before the OGLE and EP detections of the outburst. Furthermore, the next ATLAS measurements, at MJD 60458.14, were at a magnitude of $\sim$14.95, significantly above the quiescent level. This implies that the outburst must have lasted some $\sim$2 days. 

As in the case of the OGLE light curve, the brightness after outburst is less than before outburst, although the difference is not as marked. This is likely due to the different effective wavelengths of the I-band filter (7000$-$9000\AA) used in OGLE and the o-band filter (5600$-$8100\AA) used in ATLAS, implying the difference is wavelength dependent.

\begin{figure}
	\includegraphics[width=8cm,angle=-0]{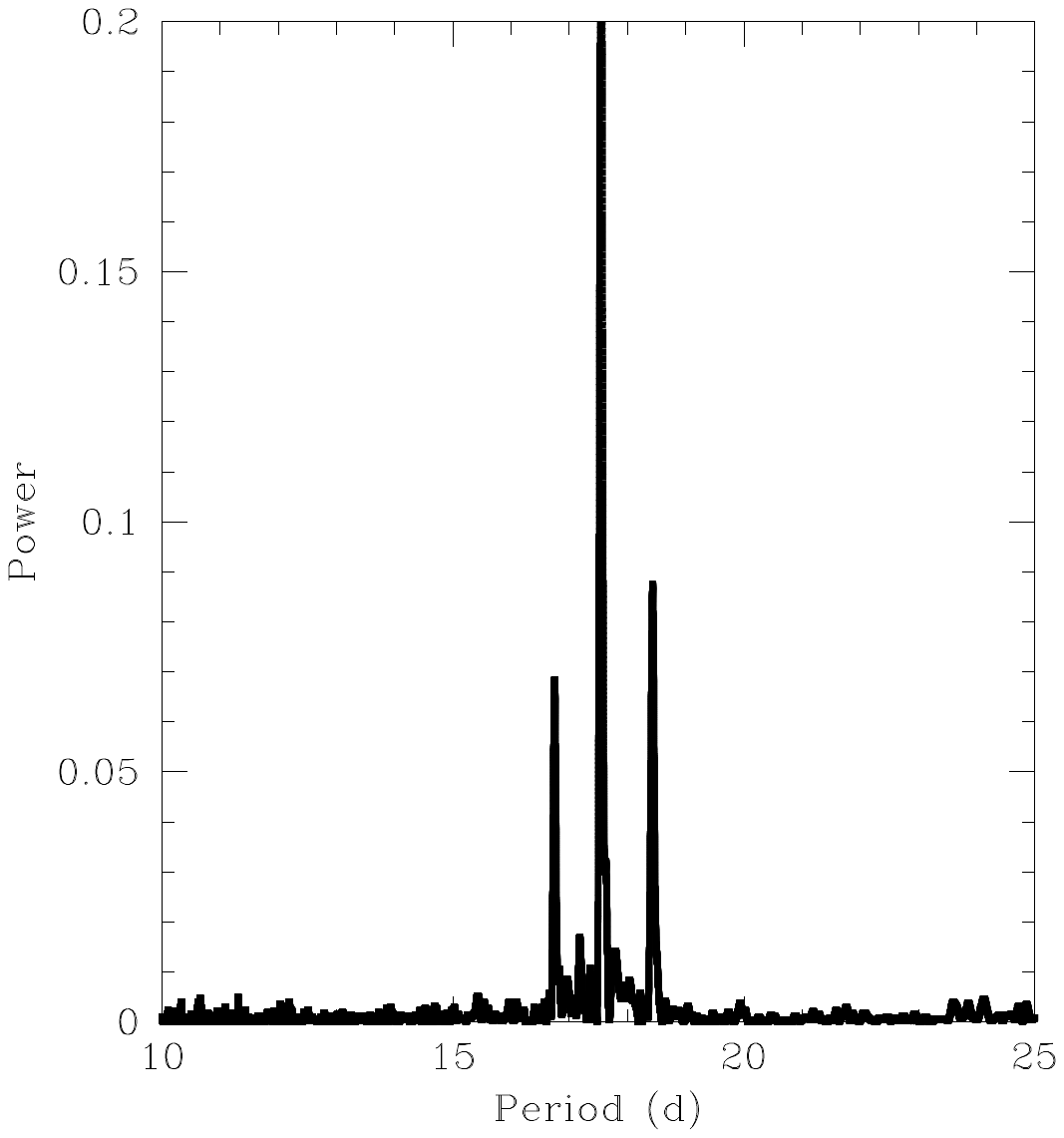}
    \caption{Lomb-Scargle power spectrum of the OGLE observations prior to JD 2458869 . The main peak is at 17.55~days, the side peaks arise from the annual sampling.}
    \label{fig:ls}
\end{figure}

\subsubsection{Period analysis}
The subset of detrended OGLE data prior to JD 2458869 were searched using a Phase Binned Analysis of Variance (AoV) method for possible periodicities in the range 2 - 100d. A clear peak was seen at a period of 17.55d - see Fig.~\ref{fig:ls}. 

\begin{figure}
	\includegraphics[width=6cm,angle=-90]{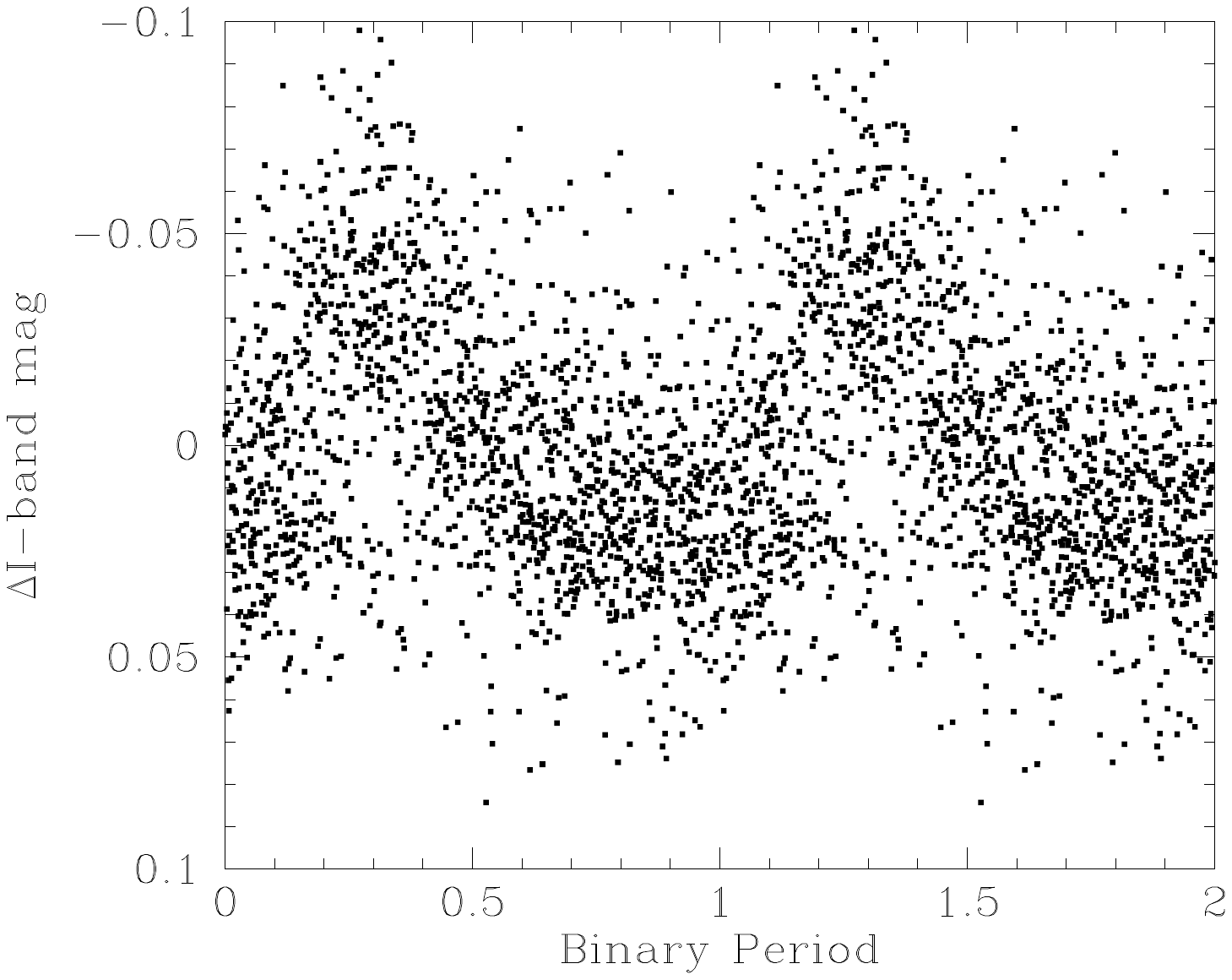}
    \caption{The detrended OGLE data prior to JD 2458869 folded at the period of 17.55~days, showing the overall sinusoidal profile and the scatter in the measurements.}
    \label{fig:fold}
\end{figure}

This period is consistent with the earlier reported period of 17.54d by \cite{2012sarraj} from analysis of MACHO data, and the more recent report of the same period by \cite{2024Treiber} using OGLE data. The OGLE data were then folded at the period of 17.55d and the resulting profile is shown in Fig.~\ref{fig:fold}. Though there is some scatter in the individual cycles, there clearly exists a dominant peak in the I-band light curve approximately at the phase $\sim$ 0.25.

Hence the OGLE data prior to JD 2458869 reveal an ephemeris for the time of the maximum of the optical profile, $T_{\mathrm{opt}}$, defined by:
\begin{equation}
T_{\mathrm{opt}} = 2456110.7\pm0.1 + N(17.5497\pm0.0008)~\textrm{JD}\label{eq:1}
\end{equation}

It is interesting to note that the phase of the outburst reported here using this ephemeris is 0.68. This is unusual as most \bexrb{} systems clearly show a strong correlation between the optical peak of any modulation and the phase of the X-ray outburst. This is believed to be because the optical peak is thought to reveal the time of periastron of the system. In this case no such correlation is evident suggesting that the X-ray outburst has arisen from a completely different trigger mechanism.

\begin{figure*}
	\includegraphics[width=15cm,angle=-0]{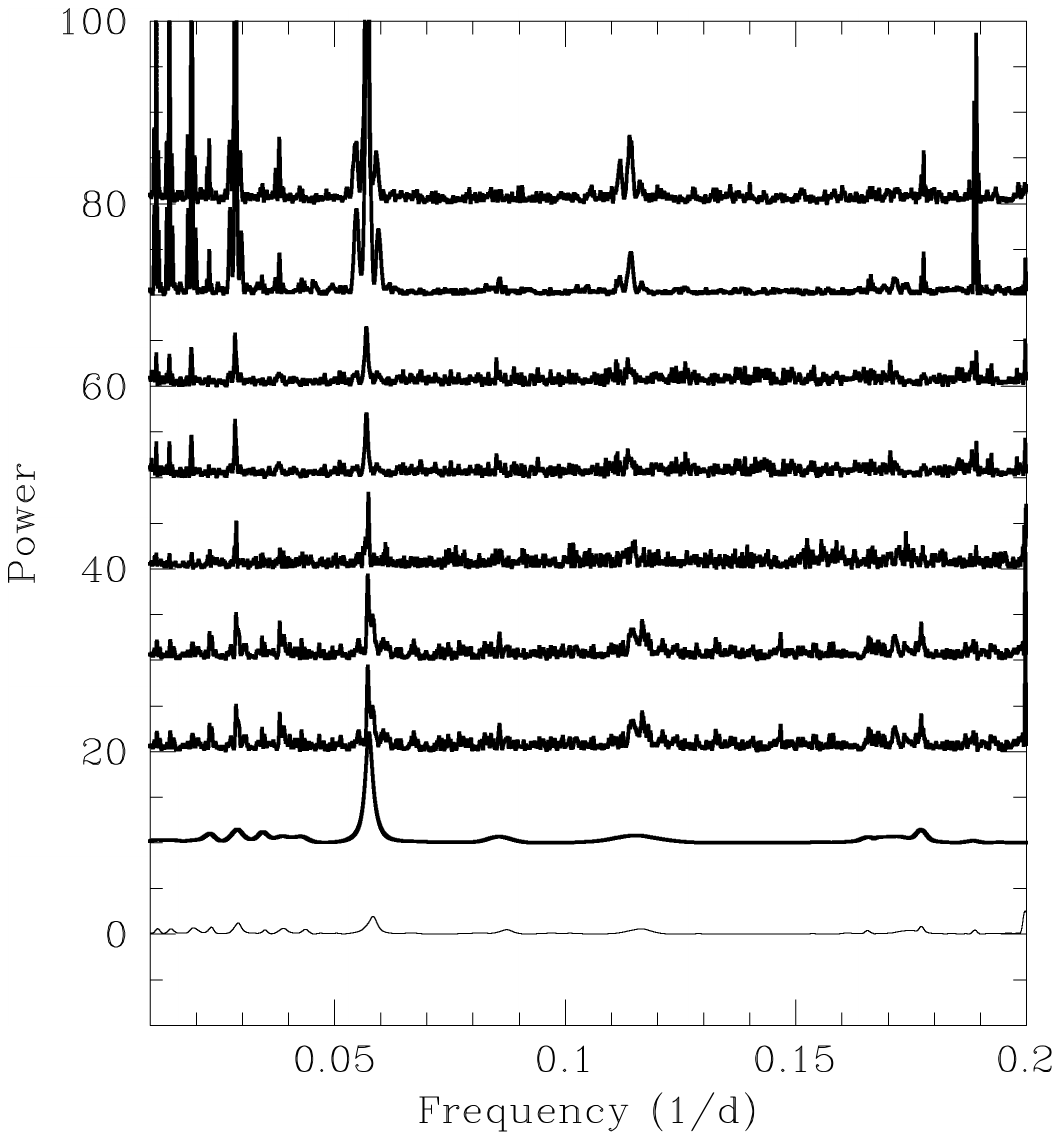}
    \caption{Time sequence of power spectra of OGLE observations divided up into 2 year groupings, except the last 2 that are 1 year samples. The earliest spectrum is at the top. The lowest 2 plots are scaled down in power by a factor of 50, but can be seen in full detail in Figure \ref{fig:lslast2}.}
    \label{fig:lsall}
\end{figure*}

\begin{figure}
	\includegraphics[width=8cm,angle=-0]{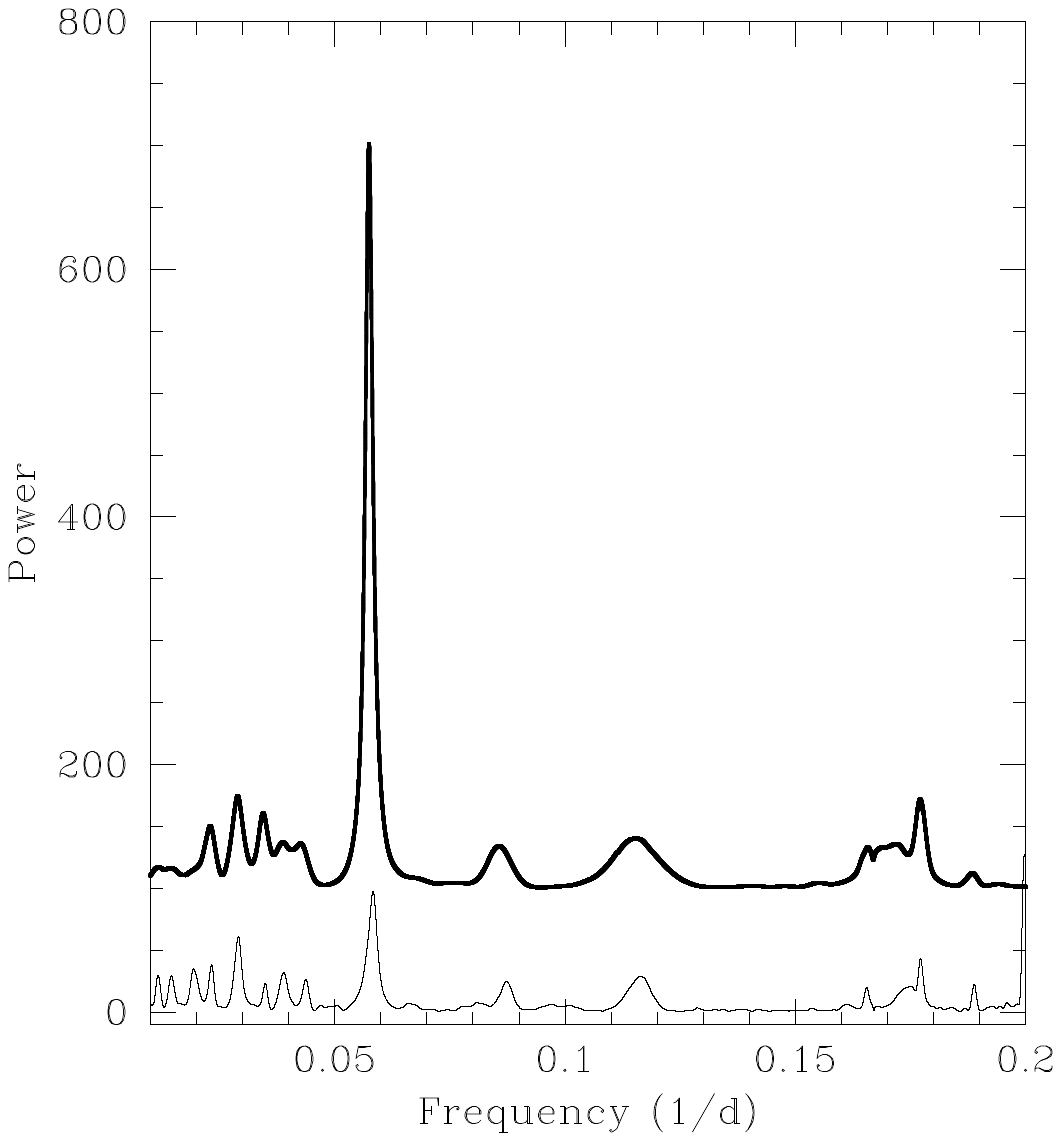}
    \caption{Power spectrum of the OGLE observations from the last two years of observations (2022 (top) and 2023 (lower)).The peaks correspond to periods of 17.41d (upper) and 17.15d (lower). Other peaks are harmonics of these periods - see text.}
    \label{fig:lslast2}
\end{figure}

Figure \ref{fig:lsall} shows power spectra of most of the OGLE data  - time goes top -> down. The first 7 are each 2 year samples (i.e. 14 years worth of data), last 2 are each 1 year and represent the most recent 2 years of observations following the COVID-19 observational gap. The last 2 are scaled down by a factor of 50 in power to fit with the others (higher power in the recent, much higher cadence observations). These last two can be seen in full detail in Figure \ref{fig:lslast2}. The dominant period, P, is clear in each case, and secondary peaks that can be seen are harmonics of the
main period: 0.029 = 2*P, 0.118 = P/2, 0.177 = P/3 and 0.086 = P * 3 / 2..

\begin{figure}
	\includegraphics[width=8cm,angle=-0]{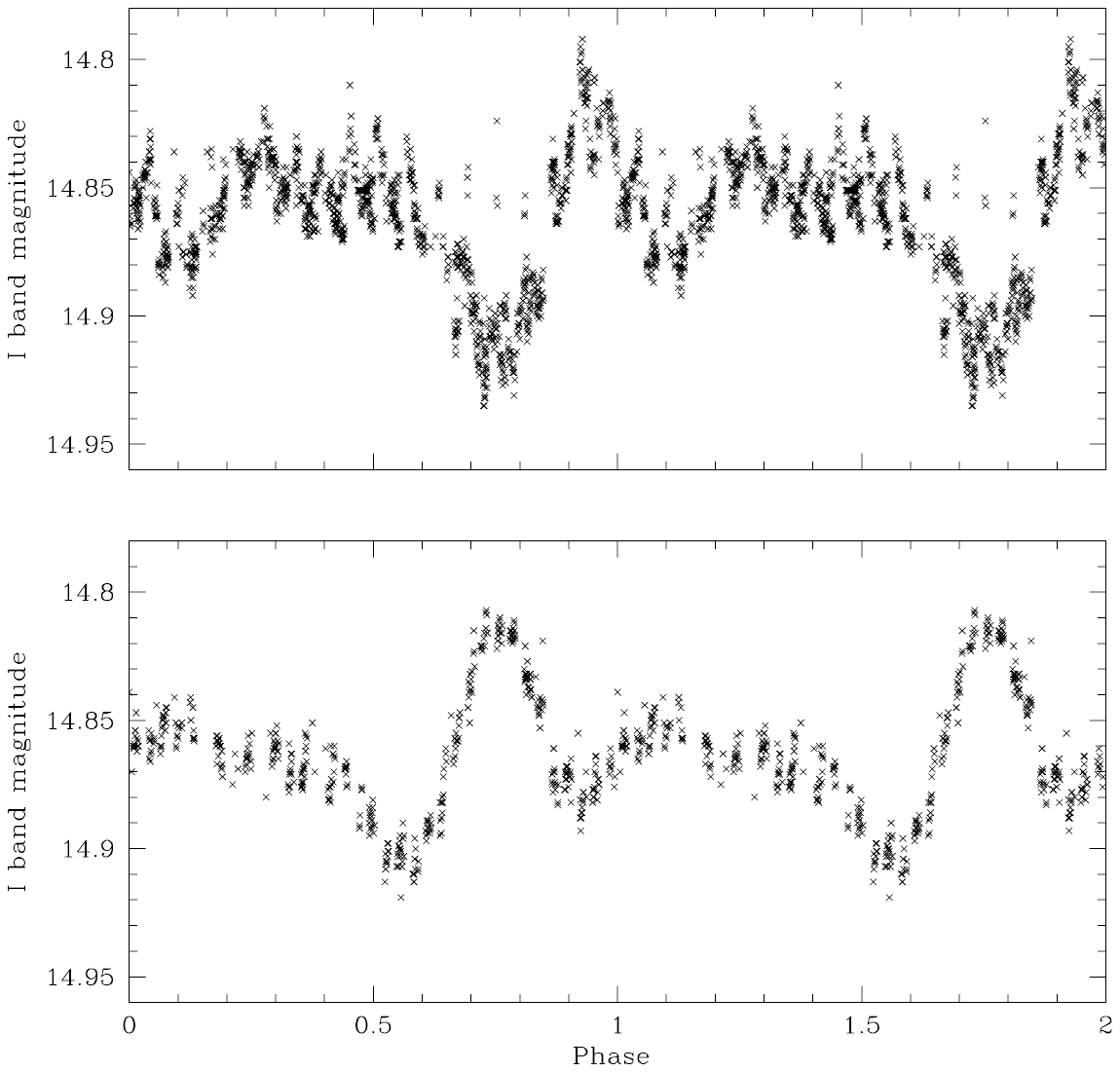}
    \caption{OGLE observations from the last 2 pre-outburst seasons detrended and folded at a periods of 17.41d (upper) and 17.15d (lower). The upper plot is based upon data from the time period JD 2459800 - 259928, the lower plot from time period JD 2460126 - 2460352. In each case the absolute phase is arbitrary.}
    \label{fig:fold2}
\end{figure}

So, from the start of MACHO observations (1992) until just before the 2 year COVID-19 gap in the OGLE data ($\sim$2020-22 or MJD 58869 - 59800), there is observed to be a persistent period of 17.55d. During this period it appears the folded profile was very sinusoidal in nature typical of most \bexrb{} systems (see, for example \cite{2023coe}), where the Be star's circumstellar disc is regularly distorted by the orbital presence of the neutron star. 
Then as we go beyond the COVID-19 gap ($\sim$2022-24 or MJD 59800 - 60352), there is evidence of a clear changes in the period. The first observing year reveals a period of 17.408 $\pm$ 0.017d, then in the folowing year the period further reduces to 17.146 $\pm$ 0.008. The folded profiles for both observing sessions are shown in Figure \ref{fig:fold2}. This figure shows the rather different folded profiles observed in these last 2 years leading up to the outburst. This difference and the period changes are discussed further in Section~\ref{discussion} below.

\subsection{SALT observations}

\begin{figure}
	\includegraphics[width=8cm,angle=-0]{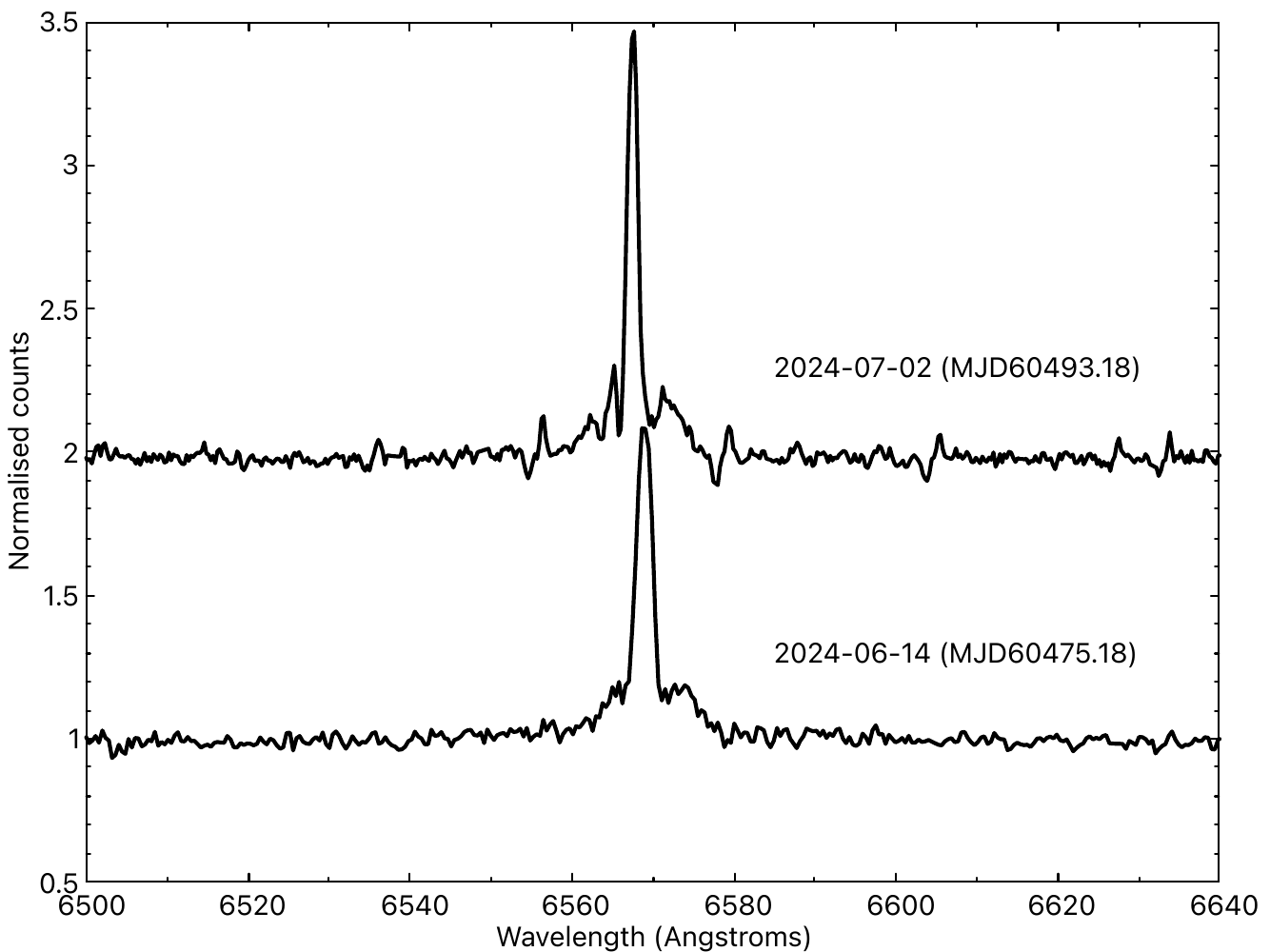}
    \caption{SALT spectra of \scwd ~covering the H$\alpha$ region.}
    \label{fig:salt}
\end{figure}

\begin{figure}
	\includegraphics[width=8cm,angle=-0]{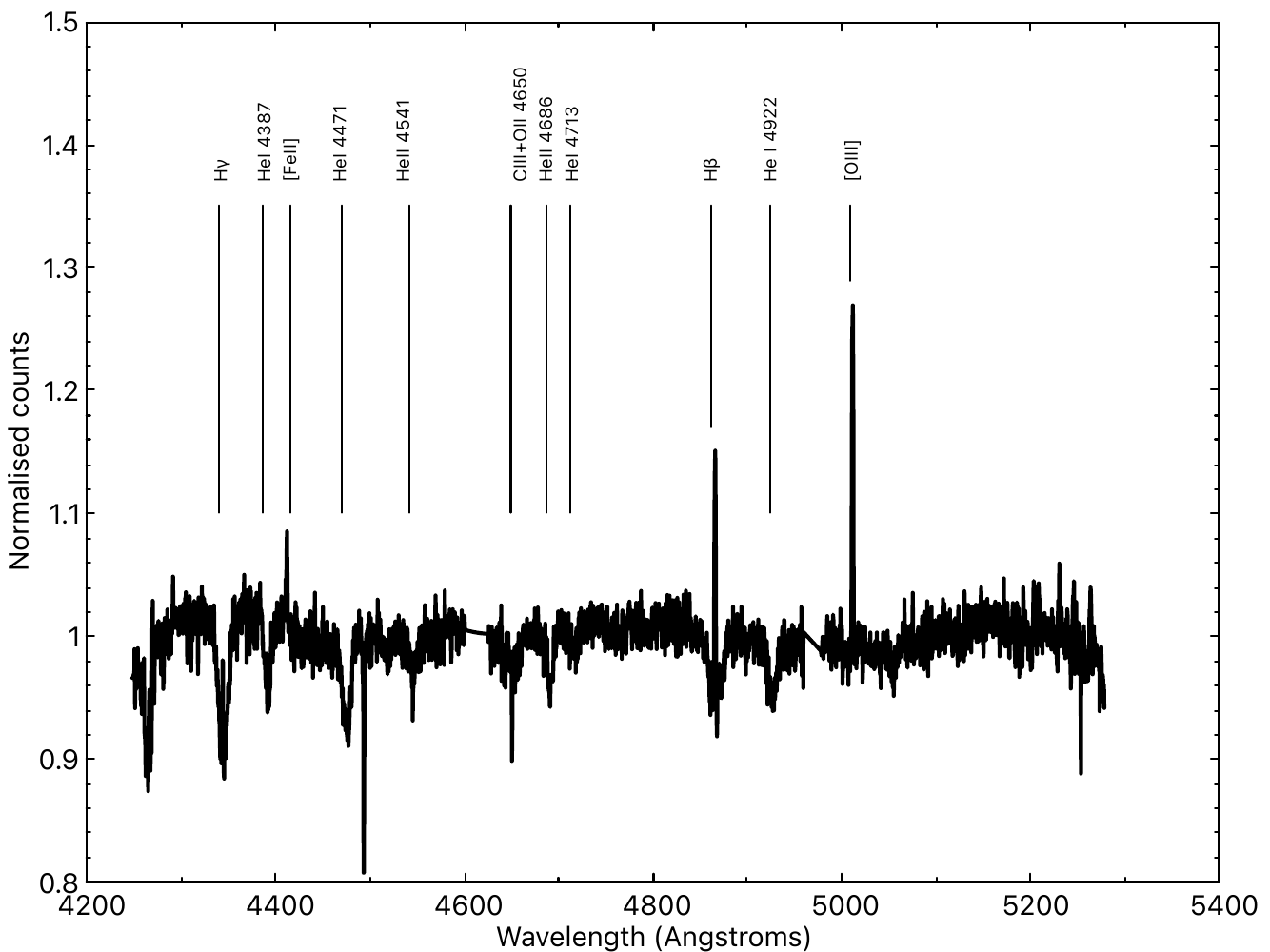}
    \caption{SALT spectrum of \scwd ~covering the blue region. Different line species are labeled at their expected rest wavelengths.}
    \label{fig:salt_blue}
\end{figure}

\begin{figure}
	\includegraphics[width=8cm,angle=-0]{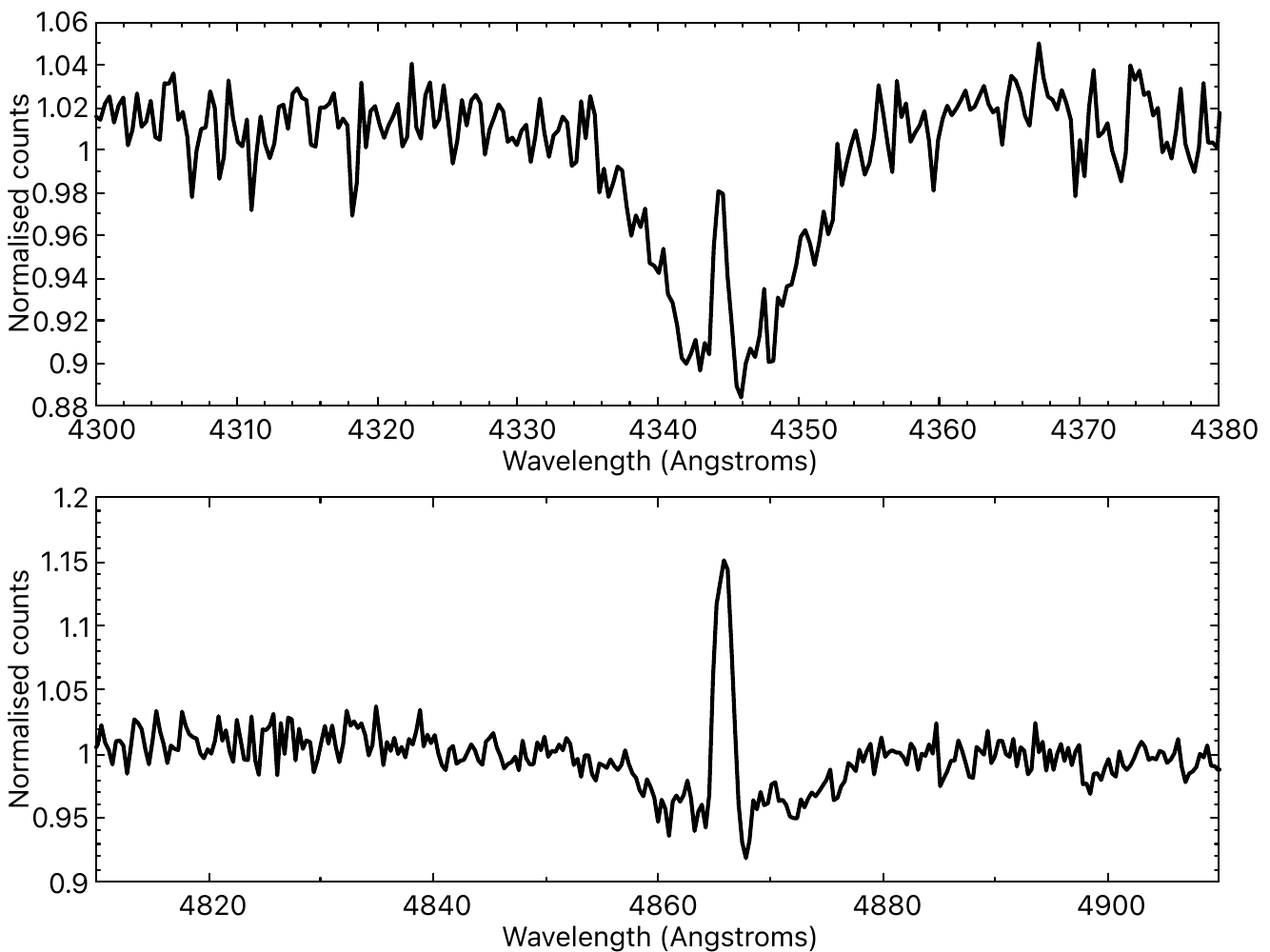}
    \caption{The H$\beta$ and H$\gamma$ absorption lines exhibiting infilling.}
    \label{fig:salt_blue_zoom}
\end{figure}

\scwd{} was observed with the Southern African Large Telescope (SALT) on 2024 June 14 (JD 2460475.7; 17.8 d after outburst) and 2024 July 02 (JD2460493.7; 35.8 d after outburst) using the Robert Stobie Spectrograph \citep{Burgh03}. The observations were carried out using several grating settings that cover different wavelength ranges: PG1800 (grating angle = 36.125$^\circ$; 5800 -- 7000~\AA), PG2300 (grating angle = 33.5$^\circ$; 4250 -- 5270~\AA) and PG2300 (grating angle = 46.25$^\circ$; 5800 -- 6660~\AA). The primary data reduction steps, including overscan correction, bias subtraction, gain
correction, and amplifier cross-talk correction, were executed using the SALT science pipeline \citep{2012ascl.soft07010C}. The remainder of the steps, consisting of arc line identification, background subtraction and 1D spectrum extraction were performed with iraf\footnote{Image Reduction and Analysis Facility: iraf.noao.edu}. The data were corrected for the heliocenter and the SMC redshift. The spectra are shown in Figs.~\ref{fig:salt} and ~\ref{fig:salt_blue}. The H$\alpha$ emission line (Fig.~\ref{fig:salt}) exhibits a strong, narrow central emission component with broad wings. The morphology of the H$\alpha$ emission line is very similar to that seen in \cite{2009Antoniou}. The narrow emission line component is possibly dominated by interstellar contamination and a circumbinary shell originating from the nova eruption.\\
The spectrum covering the blue wavelength range shows several Balmer and helium lines typical of early-type stars. The Balmer lines in this region show strong infilling, likely from the Be circumstellar disc emission and the circumbinary/interstellar emission (Fig.~\ref{fig:salt_blue_zoom}). So the broad Be-disk emission fills in the B star's photospheric absorption lines, while the circumbinary emission results in narrow emission cores. A blue-end spectrum was used to perform a spectral classification of the massive companion using the criteria in \cite{Evans2004}. The presence of the HeII4541 and HeII4686 lines constrains the spectral type to earlier than B0.5 while the ratio HeI4387/HeII4541 > 1 suggests a spectral type later than O8.5. Our spectrum does not reach far enough blue to use other helium absorption line indicators to further distinguish between O9 and B0 spectral types (e.g. HeI4143 and HeII4200). Using the distance modulus of the SMC of 18.977 \citep{2020Graczyk} and a V-band magnitude of 14.936 for \scwd{} \citep{Antoniou2019}, this constrains the luminosity class to V \citep{Straizys1981,Pecaut2013}. In summary, the SALT spectra suggest a spectral class of O9-B0Ve for the massive companion, in agreement with \cite{2009Antoniou}. It is noted that the emission lines of [FeII]4415 and [OIII]5007 that are present are typically associated with interstellar contamination and nova shells (e.g. \citealt{Mondal2018}). 

\section{Discussion}\label{discussion}

\subsection{The X-ray outburst origin}
\begin{figure}
	\includegraphics[width=9cm,angle=-0]{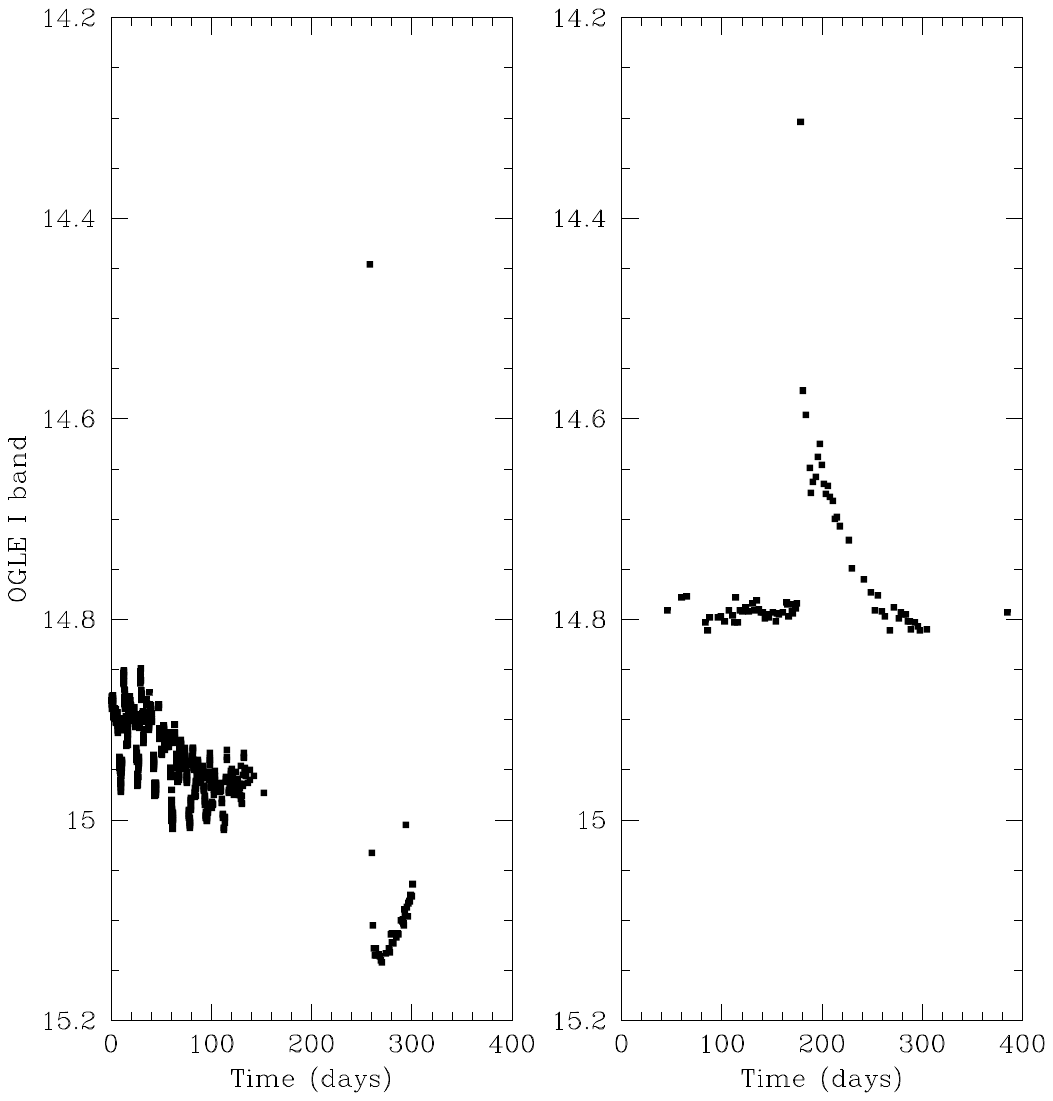}
    \caption{Comparison of the OGLE I-band outburst data for \scwd{} (left panel) and MAXI 0158-744 (right panel).}
    \label{fig:2bursts}
\end{figure}

The X-ray spectrum of \scwd{} during outburst clearly indicates that this system contains a WD super soft X-ray source with a mass of $M_{WD} \approx 1.2$ M$_\odot$. The detection of a 0.4 keV CVI edge and a 0.871 keV OVIII edge indicate that this WD is likely a massive CO WD at the upper limit of its mass range. While the X-ray spectrum of this source is similar to other known BeWD systems that have been found in the Magellanic Clouds \citep{coe2020b, 2021Kennea}, the large X-ray luminosity and short duration of this event are not typical of BeWD systems. A high-luminosity, short duration X-ray outburst has only been observed in a BeWD system during the ultraluminous nova produced by MAXI J0158-477. This outburst appears to be similar to that bright event. As seen in Figure \ref{fig:2bursts}, the initial OGLE brightening in both \scwd{} and MAXI J0158-477 both show a sudden increase in brightness by 0.5 magnitudes before rapidly fading away. In MAXI J0158-477, the source faded halfway back to quiescence on rapid timescales before dimming back to the quiescent magnitude gradually over approximately 100 days. Conversely \scwd{} fades back to the pre-outburst luminosity within a day. 

X-ray data also shows similarities between the two eruptions. X-ray emission from MAXI J0158-477 was reported to be below the detection threshold of Swift within 2 weeks after gradual dimming. Due to the observation constraints of Swift, we cannot say for certain when the outburst of \scwd{} faded beyond detection. The source was observable by NICER through at least 2024 May 29 \citep{2024Jaisawal}, but was undetectable by Swift on 2024 June 12, which places an upper limit of 16 days on the duration of the outburst. The main difference between the two events is the peak luminosity that was reached by each outburst. MAXI J0185-477 reached a maximum luminosity of $L_X = 2.3 \times 10^{40}$ erg s$^{-1}$, which is more than an order of magnitude higher than the luminosity observed in the outburst of \scwd{}. The super-Eddington luminosity combined with the brief duration of these outbursts make them hard to explain. \citet{2012Li} attempted to explain the MAXI J0158-477 outburst event as the typical shock-heating of plasma that is expected to follow a nova eruption. However, \citet{2013Morii} found that the rapid onset of super-soft X-ray emission and the cool plasma temperature of the blackbody radiation were difficult to explain via shock heating. Instead, the preferred explanation was the onset of the initial thermonuclear runaway (TNR) phase of the nova eruption with an abnormally small ejecta mass. Additionally, this source was found to contain an extraordinarily-massive WD with a mass that was near the Chandrasekhar limit. A plausible scenario that could explain this new outburst of \scwd{} is that a less-massive 1.2 M$_\odot$ WD produced a similar event. In this scenario, the X-ray flash is observed to be the initial TNR phase of a nova outburst on the surface of a massive WD. The rapid decay time indicates that very little mass was ejected during the outburst. 

An alternative model that could explain this outburst is the presence of a localized thermonuclear runaway effect \citep{1982Shara} that is sometimes referred to as a micronova \citep{2022Scaringi}. Micronovae are short-duration outbursts that last for less than a day and release up to $\sim$10$^{38}$-10$^{39}$ erg of energy \citep{2022Scaringi, 2022Schaefer}. Recent models \citep{2022Scaringi, 2022ScaringiB} identify this type of outburst as a TNR phenomenon where the material that is confined to the polar accretion cap of a magnetic WD is ignited, producing the outburst. This type of outburst is thought to be similar to a Type I X-ray outburst in BeXRB systems containing a NS \citep{1986Stella}, as the outburst lasts for a much shorter time and is less energetic than a nova by $\sim$10$^{6}$ erg \citep{2022ScaringiB}. 

The brief several hour duration of the optical outburst and large X-ray luminosity produced by \scwd{} are in excellent agreement with the properties of micronovae. However, the emission region of the source is much too large to be associated with just the polar accretion cap of a magnetized WD. The model of \citet{2022ScaringiB} finds that the fractional accretion area of a 1.2 M$_{\odot}$ WD is $f \approx 10^{-4}$. The best-fitting maximum emission region for the XRT spectrum of \scwd{} taken during outburst is $11,684 \pm 64$ km, which is $\approx$2-3 times larger than the radius of a 1.2 M$_{\odot}$ WD. This emission region indicates that emission is occurring over the entire surface of the WD instead of just at the polar caps which would be a feature characteristic to a nova eruption. Additionally, there have been no previously-observed eruptions in the \scwd{} system despite recurrent eruptions being a distinctive feature of micronova-producing systems (e.g. TV Col; \citealt{1984Szkody}). Therefore, we must conclude that a very-fast nova is preferred as an explanation for this rapid X-ray outburst.

\subsection{Be disc size}
H$\alpha$ equivalent width (EW) measurements have been shown to correlate with radii measurements from optical interferometry of isolated Be stars \citep{Grundstrom2006}. The measured EW can be used to estimate the physical size of the H$\alpha$ emitting region. \cite{Hanuschik1989} demonstrated that the peak separation of the H$\alpha$ emission line, $\Delta V$, is related to the EW by the expression:
\begin{equation}
\log \left( \frac{\Delta V}{2v\sin i} \right) = -0.32\log (-EW) - 0.20,
\end{equation}
where $v\sin i$ is the projected rotational velocity of the star. Similar relationships have been derived with slightly different coefficients (e.g. \citealt{Zamanov2016, Monageng2017}). \cite{Huang1972} showed that in a Keplerian disc, the following relation holds:
\begin{equation}
\frac{\Delta V}{2v\sin i} = r^{1/2}_{\textrm{disc}}.
\end{equation}

The total EW of the H$\alpha$ line from our SALT observations measures at $-5.41\pm0.12$~\AA~(MJD60493.18) ~and $-5.52\pm0.35$~\AA~(MJD60475.18). These two measurements are in agreement, to within errors, with the measurement of $-5.38\pm0.11$~\AA~from \cite{2009Antoniou}. To isolate the contribution of the Be disc to the H$\alpha$ emission, we fitted a Gaussian function to the narrow central component arising from the interstellar emission/circumbinary disc and subtracted its EW measurement from the total measurement. This results in EW measurements of $-2.64\pm0.12$~\AA~(MJD60493.18) ~and $-3.06\pm0.13$~\AA~(MJD60475.18) from the broad Be disc contribution. Using the two equations above and the measured EWs, the Be disc radius estimates are $r\sim 4.7$~R$_\ast$ and $r\sim 5.1$~R$_\ast$, respectively. For the spectral type of O9.5IIIe, this translates to $r\sim 58$~R$_\odot$ and $r\sim 63$~R$_\odot$ ($R_\ast \sim 12.3$~R$_\odot$; \citealt{Straizys1981}). From the derived orbital period of 17.55~d and the masses of the two components of the binary, $M_\ast \sim 31$~M$\odot$ (O9IIIe spectral type; \citealt{Straizys1981}) and $M_\textrm{WD} \sim 1.1$~M$_\odot$, the semi-major axis of a circular orbit is estimated to be $ 90$~R$_\odot$ from Kepler's Laws. We note that the Be disc EW measurements are probably underestimates as some contribution from the Be disc is expected in the central core of the emission line, and so the disc may have extended to a size that is comparable to the orbit of the white dwarf.

\subsection{Colour-magnitude variations}

\begin{figure}
	\includegraphics[width=8cm,angle=-0]{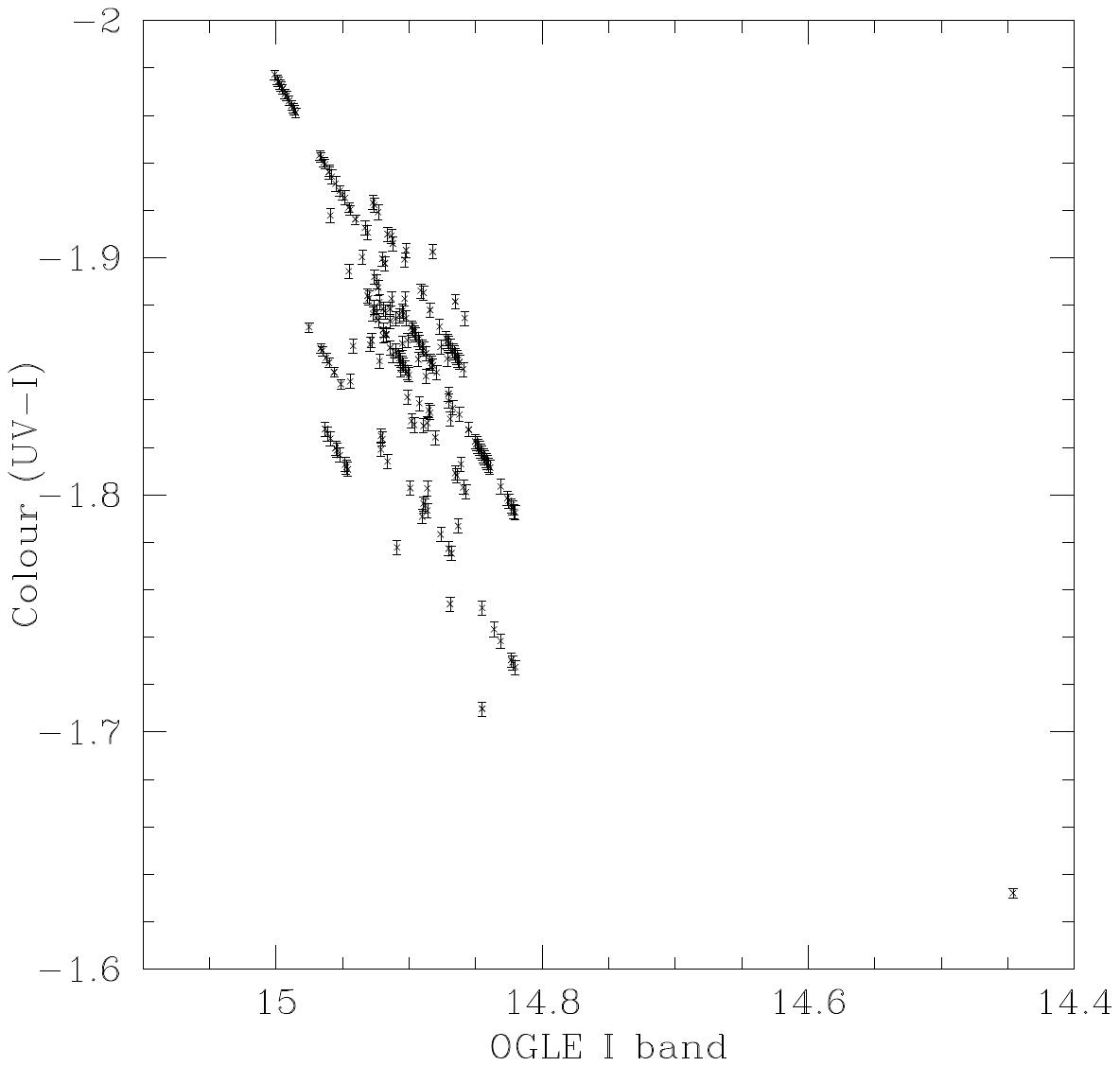}
    \caption{Colour-magnitude diagram using the UVOT and OGLE I band data. Note the one data point coincident with the source outburst in the lower right corner of the plot.}
    \label{fig:cmd}
\end{figure}

By combining the UVOT data with the I-band data it is possible to create a colour-magnitude diagram (CMD) which reveals how the overall colour of the system changed over time. The result is shown in Figure ~\ref{fig:cmd}, where the data points shown are the result of all occasions when there is a UVOT observation and an I-band observation within 3 days of each other. Since it is extremely likely that the intrinsic colours of the Be star do not vary significantly, all the variations seen in this plot arise from material outside the star - primarily, and perhaps exclusively, the circumstellar disc. Though there is little change in the brightness of the system most of the time ($\le$0.2 mag) the pattern shown clearly indicates an increased redness in the system as the I-band brightened. This is to be expected if the increasing brightness is to be attributed to an increasing size of the circumstellar disc which is, in general, cooler than the Be star. In addition, this general trend of a redder-when-brighter pattern, is believed to be indicative of inclination angles less than
$90^\circ$ \citep{1983HvaOB...7...55H, Rajoelimanana2011}.

The outstanding point in Figure ~\ref{fig:cmd} is in the lower right hand corner of the plot. This point arises from data collected at, or very close to, the outburst seen in all wavebands. It is by far the reddest state seen in the system over the previous 8 years. It is very unlikely that this rapid colour change could be due to a suddenly increasing circumstellar disc. The timescales for changes in such disks are on viscous timescales, typically 100s of days \citep{2011carciofi}. This point of view is further supported by the extremely rapid decline in the UV and optical bands immediately after the outburst on a timescale of a few days. The U(V-I) colour dropped from the value of -1.63 in the outburst back to -2.0 within 24 hours. Again, no circumstellar disc could lose the bulk of its material so very rapidly, so it is further support for the concept of a rapid catastrophic event.

Figure \ref{fig:fold2} shows the last two years of OGLE observations each folded at the determined period for each year.
It is interesting to compare these more complex recent profiles to Fig.~\ref{fig:fold} which essentially represents simple sinusoidal modulation. In these last 2 years, prior to the outburst, the change in the folded profile strongly suggests the presence of a significantly large extra structure in the circumstellar disc. This unusual component of the material surrounding the Be star, not seen in the previous 2 decades, could have played a major role in the transfer of material onto the white dwarf, and hence was integral in the triggering the bright outburst seen at all wavelengths. Detailed modelling of possible mass flow dynamics from the Be star to the white dwarf is needed to explore such a possible scenario. One promising technique would be using a Smooth Particle Hydrodynamic approach such as has been explored for accretion from Be stars on to neutron stars \citep{2019brown, 2001negueruela}.

\subsection{Superhumps?}

The change of period seen in OGLE is certainly real and it is not related to the
different cadence of the observations after 2022. For example, other
stars observed in the same programme (like pulsating or eclipsing ones) with constant period, show
exactly the same light curves in pre and post COVID-19 observations.

The shape of pulses in the \scwd ~has certainly changed. Up to JD 2456000
the pulses were quite sinusoidal and the period was ~17.54 day. In the
last two pre-COVID-19 seasons - 2018/19 (MJD 58284 - 58516) and 2019/20 (MJD 58643 - 58869) - the signal was weaker but
detectable with P=17.49 day, so essentially consistent with previous measurements. Also the period seen in the earlier MACHO
data was similar \citep{2012sarraj}.

After the COVID-19 break - in the seasons 2022/23 (MJD 59800 - 59928) and 2023/2024 (MJD 60110 - 60352) - the shape observed was
completely different - with an extra sharp peak - see Figure \ref{fig:fold2}. This shape is also clearly
seen at the very end of the 2022/23 season when the cadence was
essentially daily. The period in 2022/2023 data was 17.41d.
In the next 2023/24 season the period
decreased further to 17.15d, whilst the non-sinusoidal shape persisted. 

This behavior is similar to the so called superhump phenomena in
cataclysmic variables. Originally, they were detected in SU UMa type of
dwarf novae - in so-called super-outbursts - brighter and longer lasting
eruptions. However, the period of the superhumps was longer than the
orbital period of the system, contrary to the situation here. Such
superhumps are called "positive" superhumps. They were also found in
other classes of dwarf novae. 

However, there are also cases of the so called "negative" superhumps
where the superhump period is a few percent shorter than the orbital
one. They were found in different classes of cataclysmic variables.
Generally, the superhump period is not constant, it changes - similar
to the changes observed here in the period seen from 2022 onwards. See \cite{2024sun} for a recent review of such phenomena.

The superhump phenomenon is explained by the precession
of a tilted accretion disk - in the case of "negative" superhumps it would be a retrograde
precession. See \cite{2015thomas} for a discussion on such phenomena. From the observations presented here the precession period, $P_{pre}$,
expected in \scwd{} would be:

$1/P_{pre} = (1/17.14 - 1/17.55)$   i.e. about two years.

It is important to note that CVs have observed periods of hours while the periods seen here are of several days.
However, \scwd{} has a circumstellar disk which could be a re-scaled
version of accretion disks in CVs, so perhaps similar phenomenon can
occur there as well.

\section{Conclusions}

In this work, we report on the observation of a rapid X-ray outburst from a BeXRB system, \scwd{}. This source was first detected via an optical outburst by ATLAS on 26 May 2024 before the detection of a highly-luminous X-ray outburst state triggered follow-up multiwavelength observations. Upon follow-up observation with Swift XRT, the X-ray spectrum for this source during outburst was found to be supersoft in nature. This spectrum was best-fit by an absorbed thermal blackbody model with CVI and OVIII absorption edges. An X-ray spectrum of this nature indicates that the compact object in this BeXRB  ystem is a WD, making it only the 7th BeWD candidate ever identified and the fifth such source to be identified in the Small Magellanic Cloud. The X-ray outburst was only observed for the first few days during its brightest state and was found to have faded completely within 16 days of its onset. Optical and Ultraviolet observations taken by OGLE, ATLAS, and Swift UVOT identify a brief optical flare in the counterpart to the X-ray outburst, lasting no more than 7 days. This optical outburst peaked at between 0.2 (UVOT) and 0.5 magnitudes (OGLE) brighter than the quiescent state of the system before entering a period of sustained dimming. Optical spectroscopy confirms the optical companion to the WD in this binary is a O9Ve-B0Ve star. 

The light curve of \scwd{} is complex, and further observations will be needed in order to fully understand this unusual system. The orbital period of \scwd{} is found via multi-year OGLE monitoring to have changed appreciably in the years immediately proceeding its recent outburst, shortening from 17.55 days to 17.14 days over the course of 2 years. This period change is not well understood, and detailed mass transfer dynamical modeling will be needed to explain this behavior. One possible explanation is the ``negative superhump" phenomenon caused by the precession of a tilted accretion disk. Additional work is needed to explain the brief, X-ray outburst that reached super-Eddington luminosities for a 1.2 M$_\odot$ WD. The best comparison for the outburst that has been observed in \scwd{} is an even more luminous outburst produced by the BeWD system MAXI J0158-477 in 2011. By comparing the two events, we conclude that the X-ray outburst observed in \scwd{} is the signature of an initial thermonuclear runaway phase produced by a nova eruption on the surface of the WD. The large luminosity and rapid decay are then thought to be the result of a low ejecta mass being produced by the nova eruption, making this source the second BeWD to produce a super-Eddington X-ray luminosity during a nova.

BeWD systems are still poorly understood. With so few candidate systems identified, each system found represents a large step towards connecting BeWDs to the broader picture of massive binary stellar evolution and to important phenomena such as the ULX and ULS phenomena. In order to further our understanding of BeWDs, more systems must be identified. S-CUBED has now played an important role in identifying three of the five SMC BeWDs via soft X-ray emission behavior. It is clear that regular monitoring of the soft X-ray sky will be crucial for both identifying more of these systems in the Magellanic Clouds and finding the first candidates our own galaxy. 

\section*{Acknowledgements}
A part of this work is based on observations made with the Southern African Large Telescope (SALT), with the Large Science Programme on transients 2021-2-LSP-001 (PI: DAHB). Polish participation in SALT is funded by grant No. MEiN 2021/WK/01. 

This work has made use of data from the Asteroid Terrestrial-impact Last Alert System (ATLAS) project. The Asteroid Terrestrial-impact Last Alert System (ATLAS) project is primarily funded to search for near earth asteroids through NASA grants NN12AR55G, 80NSSC18K0284, and 80NSSC18K1575; byproducts of the NEO search include images and catalogs from the survey area. This work was partially funded by Kepler/K2 grant J1944/80NSSC19K0112 and HST GO-15889, and STFC grants ST/T000198/1 and ST/S006109/1. The ATLAS science products have been made possible through the contributions of the University of Hawaii Institute for Astronomy, the Queen’s University Belfast, the Space Telescope Science Institute, the South African Astronomical Observatory, and The Millennium Institute of Astrophysics (MAS), Chile.

This work made use of data supplied by the UK Swift Science Data Centre at the University of Leicester. J.A.K. and T.M.G. acknowledge the support of NASA contract NAS5-00136. We acknowledge the use of public data from the Swift data archive.

\section*{Data Availability}

The data underlying this article will be shared on reasonable request to the corresponding author.



\bibliographystyle{mnras}
\bibliography{sc1920} 








\bsp	
\label{lastpage}
\end{document}